\documentclass[twocolumn,superscriptaddress,floatfix,showpacs,longbibliography]{revtex4-2}

\usepackage[sc,osf]{mathpazo}
\usepackage[scaled=0.86]{berasans}  
\usepackage[colorlinks=true, allcolors=blue, urlcolor=blue]{hyperref}  
\usepackage{graphicx} 
\usepackage{amsmath,mathtools,amssymb,amsthm,bm,amsfonts,mathrsfs,bbm} 

\usepackage{xspace}  
\usepackage{pgfplots}
\usepackage{xcolor,colortbl}
\usepackage{array}
\usepackage{bigstrut}
\usepackage{mathrsfs}
\usepackage{dsfont}
\usepackage{multirow}

\usepackage{float}
\makeatletter
\let\newfloat\newfloat@ltx
\makeatother

\usepackage{algorithm}
\usepackage{algpseudocode}
\usepackage{qcircuit}

\usepackage{caption}
\usepackage{subcaption}
\usepackage{ragged2e}
\DeclareCaptionJustification{justified}{\justifying}
\usepackage{tabularx}
\newcolumntype{C}{>{\centering\arraybackslash}X}

\newcommand{\ket}[1]{| #1 \rangle}
\newcommand{\bra}[1]{\langle #1|}

\usepackage{lipsum}
\usepackage[normalem]{ulem}
\usepackage{verbatim}

\usepackage{comment}

\newcommand{\tr}{\text{tr}}

\newcommand{\be}{\begin{equation}}
\newcommand{\ee}{\end{equation}}
\newcommand{\bea}{\begin{eqnarray}}
\newcommand{\eea}{\end{eqnarray}}
\newcommand{\bes}{\begin{equation*}}
\newcommand{\ees}{\end{equation*}}
\newcommand{\beas}{\begin{eqnarray*}}
\newcommand{\eeas}{\end{eqnarray*}}

\renewcommand{\ket}[1]{|#1\rangle}
\newcommand{\ketbra}[1]{\ket{#1}\!\bra{#1}}
\renewcommand{\bra}[1]{\langle#1|}

\usepackage{xcolor}

\def\tr{\mathrm{tr}}

\begin{document}



\title{Measurement circuit ansatz: Naimark versus quantum neural-network measurements }


\author{ Sung Won Yun   }
\affiliation{Information \& Electronics Research Institute, Korea Advanced Institute of Science and Technology (KAIST), 291 Daehak-ro, Yuseong-gu, Daejeon 34141, Republic of Korea} 

\author{ Thi Ha Kyaw }
\affiliation{  LG Electronics Toronto AI Lab, Toronto, Ontario M5V 1M3, Canada }

\author{Joonwoo Bae}
\affiliation{School of Electrical Engineering, Korea Advanced Institute of Science and Technology (KAIST), 291 Daehak-ro, Yuseong-gu, Daejeon 34141, Republic of Korea }


\begin{abstract}
In this work, we present constructions of quantum circuits to implement general measurements on quantum hardware. Firstly, we investigate a quantum circuit ansatz by following the Naimark extension with a universal set of gates, such as controlled-NOT and single-qubit gates; we call it a {\it Naimark measurement}. We present a circuit ansatz framed by the Naimark extension, leaving single-qubit gates with parameters, and apply a classical optimizer to determine their parameters to approximate a desired quantum measurement. Secondly, we relax the Naimark measurement with quantum neural-network (QNN) circuits, employing parameterized quantum circuits. We present hybrid Naimark-QNN measurements by incorporating QNN circuits into Naimark measurements. Thirdly, we also consider fully QNN measurements with shallow parameterized circuits. Then, we compare the constructed measurement circuits, Naimark, hybrid Naimark-QNN, and fully QNN measurements, for strategies of state discrimination, such as minimum-error and maximum-confidence measurements. We demonstrate that QNN circuits can efficiently and effectively achieve near-optimal quantum measurements with fewer training iterations. 
\end{abstract}

\maketitle

\section{Introduction}

Measurement readout is an essential building block in quantum information processing. In particular, it is of fundamental interest to construct various forms of a quantum measurement, formulated as a positive-operator-valued measure (POVM), which can be used to reveal nonclassical effects such as hidden quantum nonlocality~\cite{PhysRevLett.74.2619, PhysRevLett.111.160402}, measurement incompatibility~\cite{Heinosaari_2016}, quantum steering~\cite{PhysRevLett.98.140402, RevModPhys.92.015001}. Quantum computing also exploits measurement readout during the computational process. For instance, mid-circuit measurements can provide outcomes that are used to update subsequent qubit operations \cite{PhysRevApplied.17.014014}. Notably, variational quantum algorithms inherently rely on measurements to solve optimization problems \cite{RevModPhys.94.015004, Cerezo:2021aa, BLEKOS20241,weber2022toward,kyaw2023boosting}.

\begin{figure}[t]
    \centering
    \includegraphics[width=0.27\textwidth]{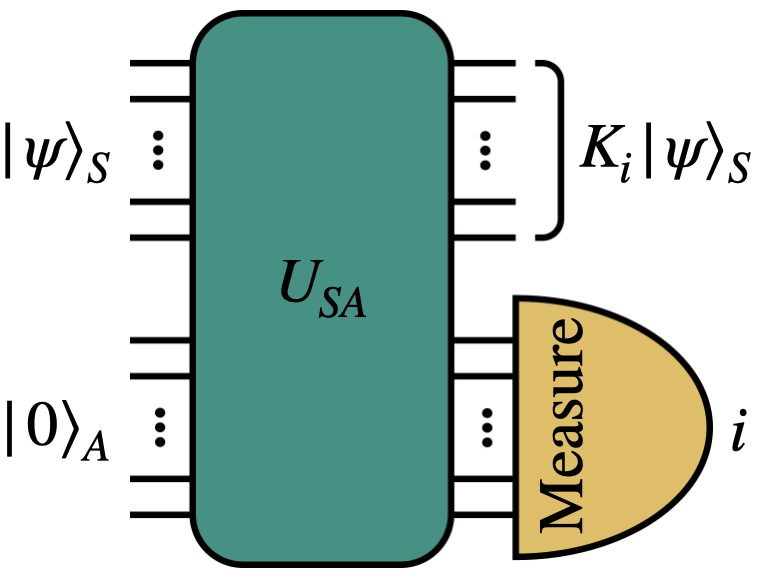}
    \caption{ The Naimark dilation theorem provides a general principle to construct an arbitrary POVM in terms of an ordinary projective measurement. The key resources are a larger Hilbert space and a quantum interaction $U_{SA}$ between the system and ancillary qubits. In a quantum circuit model, the extension to a larger Hilbert space can be realized by including a larger number of qubits. Then, a quantum circuit can implement a general measurement via the Naimark extension. A circuit $U_{SA}$ over a system and ancilla qubits is performed, and an outcome on the ancilla qubits, say $i$ (yellow), implements a general operation described by a Kraus operator $K_i$, where $K_i = _{A}\langle i | U_{SA}|0\rangle_A $, see the derivation in Eq. (\ref{eq:de}).}
    \label{fig: scheme}
\end{figure}

Implementations of POVMs can be generally obtained by referring to the Naimark extension~\cite{Naimark1943}, see also Fig.~\ref {fig: scheme}. 

\bigskip
 {\it The Naimark dilation theorem} (informal statement) \cite{Naimark1943}: Every generalized quantum measurement represented by a POVM can be implemented as an ordinary projective measurement by embedding the system into a larger Hilbert space. 
\bigskip

Applying the theorem to qubits for quantum information processing realizes a general quantum dynamics by applying a unitary transformation and measuring ancillary qubits. For a state of a system, denoted by $|\psi\rangle$, its general dynamics can be described by a unitary interaction between the system and $m$ ancillary qubits as follows, 
\bea
U_{SA} |\psi \rangle_S |0^{ } \rangle_{A}^{\otimes m} = 
\sum_{i=1}^l \left[ K_{i} |\psi_{} \rangle_S \right] | i \rangle_A \label{eq:kraus}
\eea
where $\{ K_i\}$ denote Kraus operators acting on systems. One can derive an explicit form of a Kraus operator as follows,
\bea
U_{SA} |\psi \rangle_S |0^{ } \rangle_{A}^{\otimes m} & = & \left(\sum_i | i\rangle_A\langle i|\right)~ U_{SA} |\psi \rangle_S |0^{ } \rangle_{A}^{\otimes m} \nonumber \\
& = & 
\sum_i \langle i| U_{SA}|0^{ } \rangle_{A}^{\otimes m} |\psi \rangle_S | i\rangle_A \label{eq:de}
\eea
where we have used the relation $\sum_i | i\rangle_A\langle i| = I_A$ and can identify $K_i = _A\langle i| U_{SA}|0^{ } \rangle_{A}^{\otimes m} $, that acts on systems only, see Fig. \ref{fig: scheme}.

Measurements on $m$ ancillary qubits in the computational basis $\{|i\rangle \langle i| \}$, where $i=i_1i_2\cdots i_m$ for $i_j \in \{0,1 \}$ for $j=1,\cdots, m$, find the probabilities $p(i) = \langle \psi | K_{i}^{\dagger}K_i  |\psi\rangle $, in which $K_{i}^{\dagger}K_i$ constructs a POVM element. Throughout, $l$-outcome POVM elements, $M = \{ M_i\}_{i=1}^{l}$, can be implemented by the Naimark extension in Eq.~(\ref{eq:kraus}) via Kraus operators $\{K_i \}$. We may also include an element $M_0$ that collects inconclusive outcomes, so that $\sum_{i=0}^l M_i = I$. 
 
In Eq. (\ref{eq:kraus}), one can observe that a circuit $U_{SA}$ constructs a Kraus operator with fixed ancillary qubits prepared in the computational basis. We call the circuit $U_{SA}$ in Eq. (\ref{eq:kraus}), which constructs a Naimark measurement, a {\it measurement circuit}. Note also that the complexity of implementing a POVM amounts to the construction of a quantum circuit $U_{SA}$ in Eq.~(\ref{eq:kraus}). Recently, a measurement circuit for a single-qubit POVM has been proposed \cite{POVM_qc_2019}.

\bigskip
 {\it Naimark measurements}: 
An arbitrary measurement described by a POVM can be implemented by a measurement circuit via the Naimark extension. The resulting POVM is called a {\it Naimark measurement}, which precisely constructs a desired POVM.

\bigskip

Note that we may structure a measurement circuit with elementary gates, such as the controlled-NOT (CNOT) and single-qubit gates~\cite{PhysRevA.52.3457}; hence, arbitrary measurements can be constructed.

One may point out that currently available quantum technologies are feasible with shallow-depth quantum circuits, but not with polynomial-depth ones, as full control over quantum errors is yet lacking~\cite{Preskill2018quantumcomputingin}. Therefore, it may be natural to seek the construction of a measurement circuit in such a way that the construction is experimentally feasible and cost-effective, possibly with minimal resources.

Here, we investigate the construction of a circuit ansatz for efficiently composing quantum measurements. The goal is to construct a measurement circuit that may be feasible on the present-day quantum hardware. To this end, we first investigate a measurement circuit for an $l$-outcome POVM on multiple qubits by following the Naimark extension, framing the circuit in terms of CNOT and single-qubit rotation gates. Then, conversely, we exploit the resulting structured frame as an ansatz on which a classical optimizer can work, and show that the building blocks of a Naimark measurement circuit introduce $ZZ$-interactions that render the classical search computationally hard. 

Then, we address the computational difficulty by replacing the $ZZ$-interacting building blocks with quantum neural network (QNN) circuits, such as parameterized quantum circuits (PQCs). We present a hybrid Naimark-QNN measurement circuit, in which only the $ZZ$-interacting blocks are replaced, and show that the number of CNOT gates is significantly reduced compared to the Naimark measurement. We also introduce a fully QNN measurement circuit consisting solely of QNN building blocks, without invoking the Naimark extension.

With measurement-circuit ansatzes, i.e., Naimark, a hybrid Naimark-QNN, and fully QNN measurements, we demonstrate the construction of measurements in the fundamental task of quantum state discrimination, considering both the minimum-error (ME) and maximum-confidence (MC) strategies. We also present a block-encoding circuit \cite{Camps_2022,Low2017} to realize MC measurements. The proof-of-principle demonstrations are performed on a real device, ibm$\_$strasbourg and the IQM Garnet, and a simulator in a noiseless setting, Qiskit Aer. We find that QNN measurement circuits are near-optimal in that they efficiently approximate Naimark measurements, while also being cost-effective: a classical optimizer works more efficiently on QNN measurements than on Naimark ones, and QNN measurements outperform Naimark ones for a smaller number of training iterations. In contrast, a Naimark measurement requires more iterations and ultimately reaches an optimal measurement, whereas its QNN counterpart does not.
In Sec.~\ref{sec:con}, we conclude with a summary of the results and remarks on QNN measurements.

This article is structured as follows. In Sec.~\ref{sec:naimark}, we investigate the Naimark construction of a measurement circuit for $l$-outcome POVMs of multiple qubits. In Sec.~\ref{sec:nnm}, we show that building blocks of a Naimark measurement contain two-qubit gates of $ZZ$-interactions, and replace them with PQCs to construct hybrid Naimark-QNN and fully QNN measurement circuits. In Sec.~\ref{sec:qsd}, we demonstrate the construction of measurement circuits for quantum state discrimination, such as ME and MC discrimination. We also present a block-encoding circuit to realize maximum-confidence measurements. In Sec.~\ref{sec:con}, we conclude the results and address remarks on NN measurements.

\section{ General measurements by the Naimark extension }
\label{sec:naimark}

In this section, we investigate the construction of a general $l$-outcome measurement on multiple qubits. We generalize the construction in Ref.~\cite{POVM_qc_2019} to multiple qubits to frame the structure of CNOT and single-qubit gates in a Naimark measurement. We present a circuit ansatz for a Naimark measurement. Note also that there have been resource-efficient constructions with a small number of ancillary qubits \cite{Andersson2008binary, Oszmaniec2017simulating, Oszmaniec2019simulating, Singal2022implementation, Reitzner:2024aa}, in which the constructions are probabilistic: a general POVM element is prepared with some probability. We here aim to construct a measurement in a deterministic manner. Let us review the single-qubit POVM in detail and then present a general quantum-circuit architecture for $l$-outcome measurements, enabling a classical optimizer to find optimal parameters for single-qubit gates.

\subsection{Single-qubit POVM} 
\label{sec:single_qubit_POVM}
The strategy of realizing a Naimark measurement on a single-qubit measurement is to sequentially apply two-qubit and single-qubit unitary gates to increase the number of outcomes~\cite{POVM_qc_2019}. To construct a two-outcome measurement, we consider a unitary transformation of a single-qubit state $|\psi\rangle$ as follows,
\bea
|\psi\rangle |0\rangle\mapsto 
(V_1 D_1 U )|\psi \rangle| 0 \rangle + 
(V_2 D_2 U ) |\psi \rangle | 1 \rangle\label{eq:2m}
\eea
where $U$ and $V_{i}$ for $i=1,2$ are single-qubit gates and $D_{i}$ are diagonal ones, 
\bea
D_1 & = & \cos\theta_1 | 0\rangle \langle 0| + \cos\theta_2 | 1\rangle \langle 1|,~\mathrm{and} \nonumber \\
D_2 & = & \sin\theta_1 | 0\rangle \langle 0| + \sin\theta_2 | 1\rangle \langle 1|. \label{eq:dd}
\eea
That is, operators $V_i D_i U$ for $i=1,2$ on a system $S$ form Kraus operators. They can be explicitly implemented in a circuit as follows,
\bea
 \Qcircuit @C=0.5em @R=0.5em {
    \lstick{} & \gate{U} & \ctrlo{1} & \ctrl{1} & \gate{V_1} & \gate{V_2} & \qw \\ 
   \lstick{} & \qw & \gate{R_Y(\theta_1)} & \gate{R_Y(\theta_2)} & \ctrlo{-1} & \ctrl{-1} & \qw
}\label{eq:2mc}
\eea
where controlled-$Y$ gates realize rotations on the $X$-$Z$ plane. Then, a measurement outcome $i$ in the second register realizes a transformation $V_iD_iU$ on a system qubit in the first register.

To generalize the construction above to $l$-outcome measurements, we need several pieces of machinery as follows. Let us consider an $L$-bit string $x = x_1 x_2\cdots x_L$, in which $x_i \in \{ 0,1\}$. Let $C$ denote a set of indices $\{ i: x_i=1 \}$ and $T$ of $\{ i: x_i=0 \}$. A collective CNOT gate on multiple qubits is defined by $x$, 
\bea
CX^{(x)}=
\big(I_C -\bigotimes_{ i\in C} \ketbra{1}_{i} \big)\bigotimes_{j \in T} I_{j} +
\bigotimes_{ i\in C} \ketbra{1}_{i} \bigotimes_{j \in T} X_{j}
\label{eq:conot}
\eea
where $T$ and $C$ are collections of target and control registers. For instance, a CNOT gate can be expressed as follows: $C^{(10)} = |0\rangle_1\langle0| \otimes I_2 +|1 \rangle_1\langle 1| \otimes X_2$.

\begin{figure*}
\centering
\includegraphics[width=1.0\textwidth]{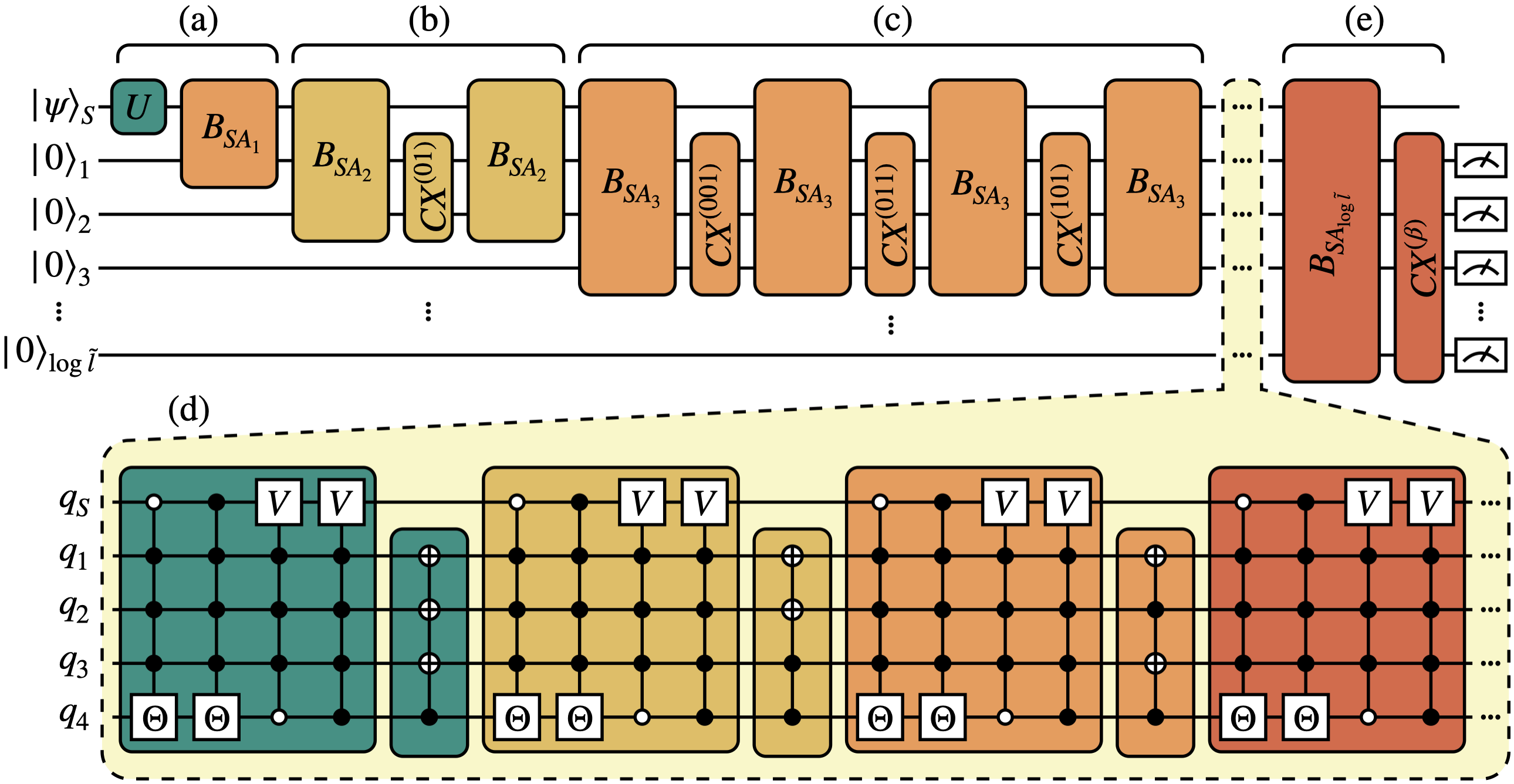} 
 \caption{The construction of a measurement circuit is detailed: $l$-outcome POVM elements for measurements on $1+\log_2 \widetilde{l}$ qubits are shown by following the Naimark theorem~\cite{POVM_qc_2019, Naimark_theorem} where $\widetilde{l}= 2^{\lceil\log_2 l\rceil}$ is the nearest power of $2$ not smaller than $l$. (a) Two-outcome measurements can be prepared after a binary module $B_{SA_1}$ in Eq.~(\ref{eq:bij}). (b) After the first $B_{SA_2}$ and $CX^{(01)}$, a three-outcome measurement is realized. To add a POVM element, gate $B_{SA_2}$ is applied, after which a four-outcome measurement is achieved. (c) Similarly, a five-outcome measurement is realized after the first $B_{SA_3}$ and $CX^{(001)}$. Then, a POVM can be added after two gates $B_{SA_3}$ and $CX$. Hence, eight-outcome measurements are realized. (d) A gate $B_{SA_4}$ is composed of CNOT gates and $Y$-rotation gate parameterized by $\Theta$. (e) shows the final gates: $B_{SA_{log_2 \widetilde{l}}}$ and $CX^{(\beta)}$. }
    \label{fig: POVMsingle}
\end{figure*}

We also introduce a binary module $B_{ij}$ for two qubits in the $i$- and $j$-th registers: the module can be described as a circuit in the following,
\bea
B_{ij} = \Qcircuit @C=0.5em @R=0.5em {
    \lstick{} &  & \ctrlo{1} & \ctrl{1} & \gate{V_1} & \gate{V_2} & \qw \\ 
    \lstick{} & \qw & \gate{R_Y(\theta_1)} & \gate{R_Y(\theta_2)} & \ctrlo{-1} & \ctrl{-1} & \qw
}. \label{eq:b12}
\eea
For a state $|\phi(i,j)\rangle= |1\rangle^{\otimes (j-i-1)}$, a module $B_{ij}$ realizes a transformation as follows, 
\bea
&& B_{ij}|\psi\rangle_i |\phi(i,j)\rangle |0\rangle_j \nonumber \\
&= &\big( K_1 |\psi\rangle_i \big)  |\phi(i,j)\rangle| 0\rangle_{j} + \big( K_2 |\psi\rangle_i \big) |\phi(i,j)\rangle |1\rangle_j \nonumber
\eea
where $K_1$ and $K_2$ are Kraus operators. Note that for $|\phi(i,j) \rangle \neq |1\rangle^{\otimes (j-i-1)}$, the binary module works as an identity, $B_{ij} = I$. The module for registers $i$ and $j$ that are not adjacent can be constructed in a circuit as follows, 
\bea
B_{ij} = \hspace{2.5em} \Qcircuit @C=0.5em @R=0.5em 
{ \lstick{i} & \qw & \ctrlo{1} & \ctrl{1} & \gate{V_1} & \gate{V_2} & \qw \\ \lstick{{i+1}} & \qw & \ctrl{1} & \ctrl{1} & \ctrl{-1} \qwx[1] & \ctrl{-1} \qwx[1]& \qw \\
&  & & &  & &  \\
&  & & \dots &  & &  \\
&  & & &  & &  \\
\lstick{{j-1}} & \qw & \ctrl{1}\qwx[-1] & \ctrl{1}\qwx[-1] & \ctrl{1}\qwx[-1] & \ctrl{1}\qwx[-1] & \qw \\ 
\lstick{j} & \qw & \gate{R_Y(\theta_1)} & \gate{R_Y(\theta_2)} & \ctrlo{-1} & \ctrl{-1} & \qw } \label{eq:bij}
\eea
where $i$, $i+1$, $j-1$, and $j$ denote registers. 


With collective CNOTs and binary modules above, we are ready to present a quantum circuit for $l$-outcome measurement on a single qubit in a constructive manner. Let us begin with two-, three-, and four-outcome POVMs. The construction of a full circuit is summarized in Algorithm \ref{alg:single_qubit_povm}, and the circuit ansatz is shown in Fig.~\ref{fig: POVMsingle}. For instance, a two-outcome measurement is shown in Fig.~\ref{fig: POVMsingle} (a): a measurement of the first ancilla register in the computational basis prepares two POVM elements in the system register $S$. 

A three-outcome POVM can be constructed by applying an additional binary module, $B_{SA_2}$ in Fig. (\ref{fig: POVMsingle}) (b). Outcomes $00$, $10$, $11$ can occur in the first and the second ancilla registers; then, three POVM elements appear in the system register, i.e.,
\bea
 K_1 |\psi\rangle_S |00\rangle_{12} + K_2 |\psi\rangle_S |10 \rangle_{12} +
 K_3 |\psi\rangle_S |11 \rangle_{12} \nonumber
\eea
from which POVM elements $M_i = K_{i}^{\dagger}K_i$ are obtained for $i=1,2,3$.

To have four outcomes in two ancilla registers, a collective CNOT gate, $CX^{(01)}$, is placed to transform $|11\rangle$ to $|01\rangle$ in the ancilla register; see Eq.~(\ref{eq:conot}), so that we have
\bea
 K_1 |\psi\rangle_S |00\rangle_{12} + K_2 |\psi\rangle_S |10 \rangle_{12} +
 K_3 |\psi\rangle_S |01 \rangle_{12}.\label{eq:3q} 
\eea
A binary module $B_{SA_2}$ creates a state $|11\rangle_{12}$; a resulting state is given by 
\bea
&& K_1 |\psi\rangle_S |00\rangle_{12} + K_2 |\psi\rangle_S |01 \rangle_{12} +\nonumber \\
&& K_3 |\psi\rangle_S |10 \rangle_{12} +K_4 |\psi\rangle_S |11\rangle _{12} \label{eq:4q} 
\eea
so that outcomes in the ancilla registers prepare four POVM elements.

To construct a five-outcome POVM, a third ancilla qubit is added. Again, one applies a binary module $B_{SA_3}$ and a collective CNOT gate ${CX^{(x)}}$. Repeating a binary module four times, we have eight-outcome POVM elements; see Fig.~\ref{fig: POVMsingle} (c). To be precise, the circuit is constructed as, $B_{S A_3} CX^{(101)}
B_{S A_3} CX^{(011)}
B_{S A_3} CX^{(001)} B_{S A_3}$, see Table~\ref{POVMsingleTable} for more outcomes.

\begin{table}[t]
\centering
\caption{ As shown in Eqs. (\ref{eq:3q}) and (\ref{eq:4q}), a POVM is implemented according to measurement outcomes on ancillary qubits. While the number of ancillary qubits, denoted by $n$, increases, so does the number of POVM elements, where each POVM element is obtained by measurement outcomes on a register of ancillary qubits. The number of POVM elements, unitary blocks, and $n$ ancillary qubits, and their states are summarized for $n=1,2,3,4,5$. The framework for an $l$-outcome POVM is constructive. }
\label{POVMsingleTable}
\begin{tabular}{c|c|c|c}
\hline
\rule{0pt}{4.8ex}
\shortstack{$\#$ of POVM \\ elements} &
\shortstack{Ancilla \\ qubits} &
\shortstack{Unitary \\ blocks} &
\shortstack{$n$-(ancillary) \\ qubit state} \\
\hline
$2$& $1$& $B_{SA_{1}}$ & $|1\rangle_1$\\
\hline
$3$& $2$&$CX^{(01)}$ $B_{SA_{2}}$ & $|01\rangle_{12}$\\
\hline
$4$& $2$& $B_{SA_{2}}$ & $|11\rangle_{12}$ \\
\hline
$5$& $3$&$CX^{(001)}$ $B_{SA_{3}}$ & $|001\rangle_{123}$\\
\hline
$6$&$3$&$CX^{(011)}$ $B_{SA_{3}}$  & $|011\rangle_{123}$ \\
\hline
$7$&$3$&$CX^{(101)}$ $B_{SA_{3}}$  & $|101\rangle_{123}$ \\
\hline
$8$&$3$&$B_{SA_{3}}$  & $|111\rangle_{123}$\\
\hline
$9$&$4$&$CX^{(0001)}$ $B_{SA_{4}}$ & $|0001\rangle_{1234}$ \\
\hline
$10$&$4$&$CX^{(0011)}$ $B_{SA_{4}}$  & $|0011\rangle_{1234}$\\
\hline
$11$&$4$&$CX^{(0101)}$ $B_{SA_{4}}$  & $|0101\rangle_{1234}$\\
\hline
$12$&$4$&$CX^{(0111)}$ $B_{SA_{4}}$ & $|0111\rangle_{1234}$\\
\hline
$13$&$4$&$CX^{(1001)}$ $B_{SA_{4}}$ & $|1001\rangle_{1234}$\\
\hline
$14$&$4$&$CX^{(1011)}$ $B_{SA_{4}}$ & $|1011\rangle_{1234}$ \\
\hline
$15$&$4$&$CX^{(1101)}$ $B_{SA_{4}}$ & $|1101\rangle_{1234}$ \\
\hline
$16$&$4$&$B_{SA_{4}}$ & $|1111\rangle_{1234}$\\
\hline
$17$&$5$&$CX^{(00001)}$ $B_{SA_{5}}$ & $|00001\rangle_{12345}$ \\
\hline
\end{tabular}
\end{table}

In general, the $i$-th $B_{SA}$ and $CX^{(x)}$ gates create $i+1$ POVM element. To this end, $L(i)$ ancilla qubits are needed, where $L(i) = \lfloor \log_2 i \rfloor +1$. Since it holds
\bea
\lfloor \log_2 i \rfloor +1 = \lceil \log_2(i+1) \rceil \nonumber
\eea
for $i\geq1$, an $l$-outcome POVM requires $\lceil \log_2l \rceil$ ancilla qubits. 

The strategy of including collective CNOT gates is as follows. Let us consider $L-1$ ancilla qubits, and we suppose that one ancilla is added, thus $L$ ancilla qubits. After the first binary module with $L$ ancilla qubits, which are denoted by
\bea
|a\rangle = |a_1 \cdots a_{L}\rangle_{L} \nonumber
\eea
one can see that $L$-bit string $a$ cannot be an element of the following set, 
\bea
A(L) &=& \{ 0_{1} 0_{2} \cdots 0_{L-2 }, 0_{L-1 }1_{L}, \nonumber \\
&& 0_{1} 0_2 \cdots 0_{L-2 } 1_{L-1 } 1_{L}, \nonumber \\
&&\cdots \nonumber \\
&& 1_{1} 1_2\cdots 1_{L-2 } 0_{L-1 } 1_{L} \}. \nonumber 
\eea
Note that the set has a cardinality $|A(L)| = 2^{L-1}-1$. This implies that one should apply a binary module $|A(L)|$ times more, in order to have $2^{L}$ POVM elements on a system after all. To this end, a collective CNOT gate should be placed between binary modules. We remark that collective CNOT gates should be ordered such that $CX^{(x)}$ with a smaller $x$ applies at an earlier time. For instance, for $L=3$ in Fig.~\ref{fig: POVMsingle} (c), collective CNOT gates are structured such that $CX^{(001)}$ applies first, $CX^{(011)}$ second, and $CX^{(101)}$ finally.

\begin{figure*}[t!]
\centering
\includegraphics[width=0.8\textwidth]{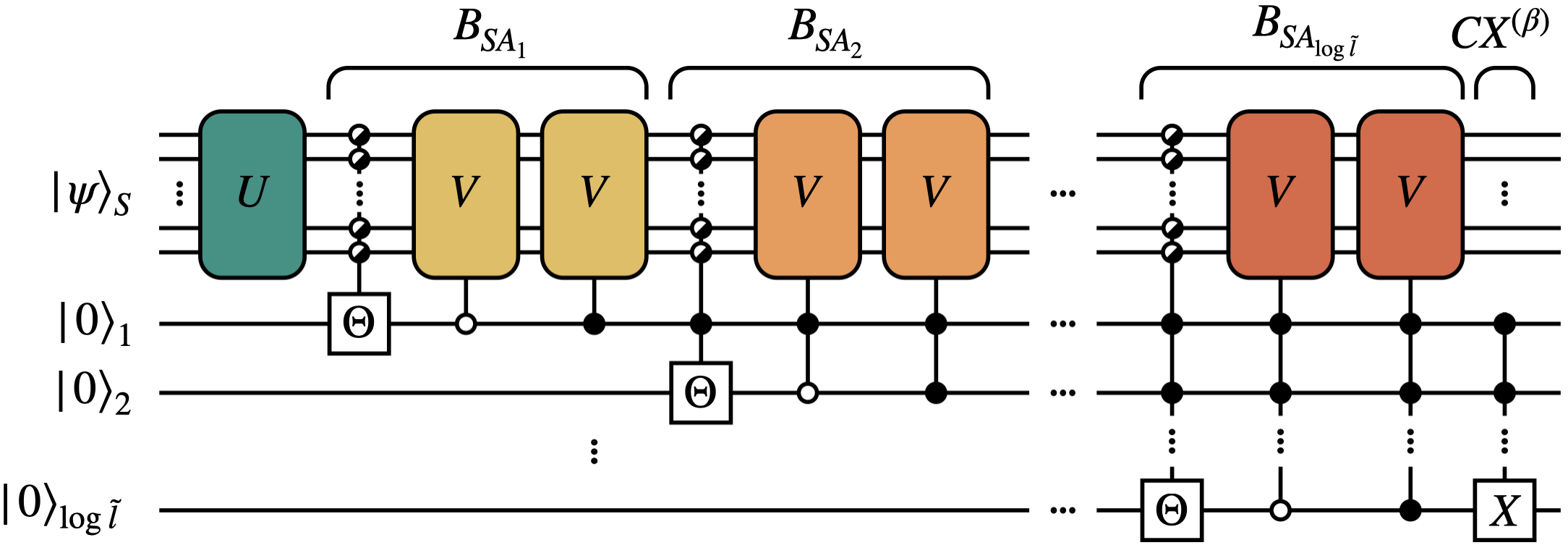} 
 \caption{Multi-qubit POVM extended from the single-qubit case. The single-qubit unitaries $U$ and $V$ are replaced with multi-qubit gates. The two single-qubit $R_Y$ rotations can be interpreted to uniformly controlled rotation gates applied to the given single-qubit state, and thus generalized to uniformly controlled rotations on the prepared multi-qubit state. The $CX^{(x)}$ gates remain unchanged. The overall construction principles of the POVM are preserved. See Fig.\ref{fig: POVMsingle} for the definitions of $\widetilde{l}$ and $\beta$.} 
    \label{fig:POVMmulti}
\end{figure*}

\subsection{Multi-qubit POVM}

\begin{figure}[h!]
\includegraphics[width=0.34 \textwidth]{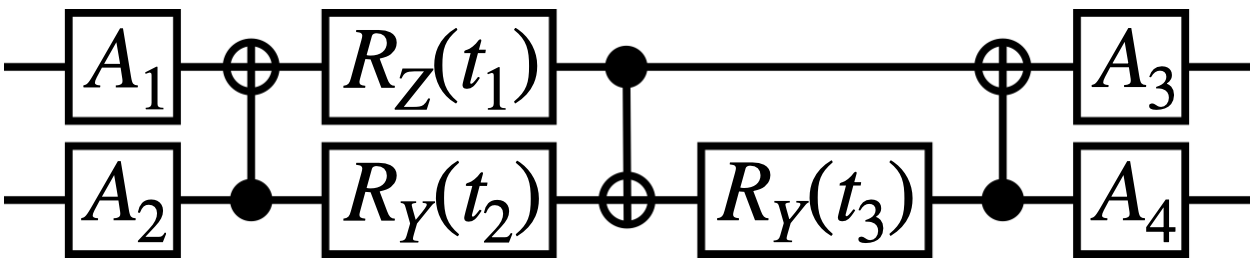}
    \centering
    \caption{ A two-qubit gate can be generally decomposed by single-qubit rotations and CNOT gates~\cite{PhysRevA.69.032315}. A canonical construction contains three CNOT gates, single-qubit rotations about $Y$ and $Z$-axis, and single-qubit gates $\{A_i\}$.}
    \label{fig:U(4)}
\end{figure}

The circuit for an $l$-outcome POVM can be extended to a multi-qubit system in general. Let us consider two-qubit POVMs. An exact construction of a two-qubit gate $U\in \textbf{U}(4)$ can be found by a canonical decomposition~\cite{PhysRevA.69.032315} 
\bea
U = (A_1 \otimes A_2) \cdot N(\alpha, \beta, \gamma) \cdot (A_3 \otimes A_4),
\eea 
where $\{ A_j \in \textbf{U}(2)\}$ are single-qubit gates and
\bea
N(\alpha, \beta, \gamma) = exp[i(\alpha \sigma_x \otimes \sigma_x + \beta \sigma_y \otimes \sigma_y + \gamma \sigma_z \otimes \sigma_z)], \nonumber
\eea 
with \(\alpha, \beta, \gamma \in \mathbb{R}\). The circuit implementation of $\textbf{U}(4)$ is shown in Fig~\ref{fig:U(4)}. Following the strategies shown in the case of a single-qubit POVM, an $l$-outcome POVM on two-qubit systems can be constructed; see Fig.~\ref{fig:POVMmulti} for a general construction.

\subsection{ Naimark measurement ansatz and the computational hardness } 

We have presented a measurement circuit for $l$-outcome measurements; see also Figs.~\ref{fig: POVMsingle} and~\ref{fig:POVMmulti}, and now exploit the circuit as an ansatz, leaving and optimizing parameters of single-qubit gates. These parameters are initialized randomly and subsequently optimized by a classical optimizer to maximize a chosen objective function.

In the optimization of parameters in a Naimark measurement ansatz, we here point out that a Naimark measurement shares the structure with quantum approximate optimization algorithms (QAOAs) for quadratic unconstrained binary optimization (QUBO), which is generally NP-Hard \cite{SciPostPhysCore.4.4.031}. Let us recall that an objective function in a QUBO problem is given as
\bea
f(x) = \sum_{j=1}^n \beta_j x_j + \sum_{i,j} \gamma_{ij}x_i x_j \nonumber
\eea
which one aims to minimize with an optimal sequence $x=x_1\cdots x_n$ where $x_i\in \{0,1\}$. It may be mapped to a cost Hamiltonian, 
\bea
H_C= \sum_{j=1}^n \beta_j {\sigma}_{j}^{(z))} + \sum_{i,j} \gamma_{ij} \beta_j {\sigma}_{i}^{(z))} {\sigma}_{j}^{(z))} \nonumber
\eea
where ${\sigma}_{j}^{(z))}$ is a Pauli $Z$-operator of the $j$-th qubit. The optimization problem can be approached by QAOAs \cite{farhi2014quantum}, which introduce an ansatz iterating a target and a mixing Hamiltonians, denoted by $H_T$ and $H_M$, respectively
\bea
U(\beta,\gamma) = \prod_{j=1}^{p} \exp[-i \beta_j H_M ] \exp[-i \gamma_j H_T ] \label{eq:qaoa} 
\eea
where the task is to optimize classical parameters $\beta = (\beta_1,\cdots, \beta_p)$ and $\gamma = (\gamma_1,\cdots, \gamma_p)$ for a value $p>0$, which is fixed {\it a priori}. Note that QAOAs achieve the optimization of a cost function when $p\rightarrow \infty$ \cite{farhi2014quantum}. Cases with finite values of $p$ are generally unexplored, although they have been investigated for some cases \cite{PhysRevLett.124.090504}.

In the ansatz in Eq. (\ref{eq:qaoa}), optimizing parameters with a target $H_T$ that consists of two-body interactions defines the main optimization problem, in which the two-body interactions are given as an Ising interaction as follows, 
\begin{equation}
\exp [ -i J_{12} \gamma_l \sigma_1^{(z)} \sigma_2^{(z)} ] = \vcenter{\Qcircuit @C=1em @R=.7em {
& \ctrl{1} & \qw & \ctrl{1} & \qw \\
& \targ & \gate{R_z(2 \gamma_l J_{12})} & \targ & \qw
}}
\label{eq:QAOA circuit}
\end{equation}
with a coupling parameter $J_{12}$, where $\gamma_l$ is a parameter in the $l$-th layer iteration in QAOAs, see Eq.~(\ref{eq:qaoa}). 

Let us revisit a Naimark measurement; see also a circuit in Eq.~(\ref{eq:2mc}), and find that a controlled rotation is decomposed as follows, 
\bea
\begin{array}{c}
    \Qcircuit @C=1em @R=1em {
        \lstick{} & \ctrl{1} & \qw \\
        \lstick{} & \gate{R_Y(\theta)} & \qw
    }
\end{array}
\ = \ 
\begin{array}{c}
    \Qcircuit @C=1em @R=1em {
        \lstick{} &\qw & \ctrl{1} & \qw & \ctrl{1} & \qw  \\
        \lstick{} & \gate{R_Y(\theta/2)} & \targ  & \gate{R_Y(-\theta/2)} &\targ & \qw
    }
\end{array}\nonumber\label{eq:r2}
\eea
which contains single-qubit rotations and CNOT gates. Note that we have used the relation $e^{i\phi Y} X = X e^{-i\phi Y}$ and also that $Y$- and $Z$-rotations are connected by elementary gates, $R_Y(\theta) = (SH) R_Z(\theta) (SH)^\dagger$ where $H$ is a Hadamard gate and $S = \mathrm{diag}[1,i]$ a phase gate. 

Hence, optimizing the parameters of a controlled rotation in a Naimark circuit reproduces, up to a local unitary, the same Ising-type structure that makes QAOA landscapes empirically difficult for local optimizers. This structural similarity suggests that a classical optimizer attempting to fix parameters of a Naimark circuit will face a landscape of local minima and slow convergence, consistent with what is observed numerically in Sec. \ref{sec:qsd}.

\begin{figure}[t]
        \centering
        \includegraphics[width=0.47 \textwidth]{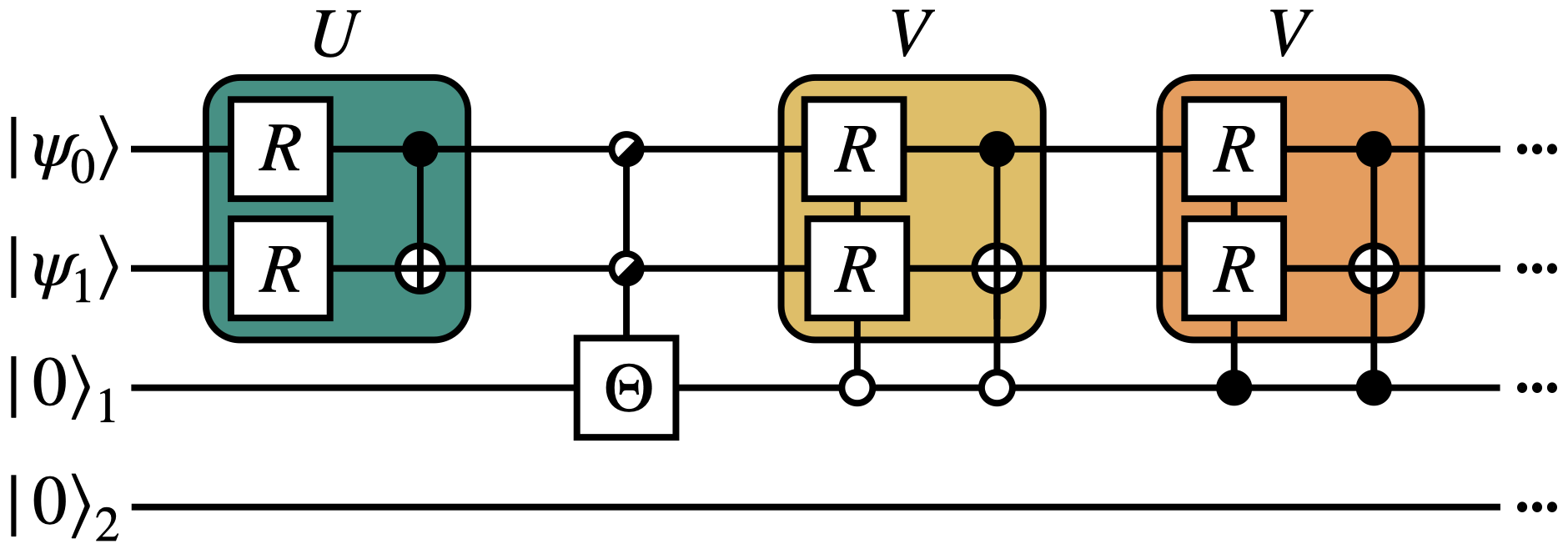}
        \caption{ A hybrid Naimark-QNN measurement for two-qubit states is shown. In a Naimark circuit in Fig.~\ref{fig:POVMmulti}, and multi-qubit unitaries on a system are replaced by PQCs. Note that a Naimark and a hybrid Naimark-QNN measurements share the same circuit structure: CNOT gates to include ancillary qubits are placed in the same way. }
            \label{fig:Yordan multi qubit PQC}
\end{figure}%

\section{{Quantum} Neural-Network Measurements}
\label{sec:nnm}

\begin{figure}[t]
        \centering
        \includegraphics[width=0.369575\textwidth]{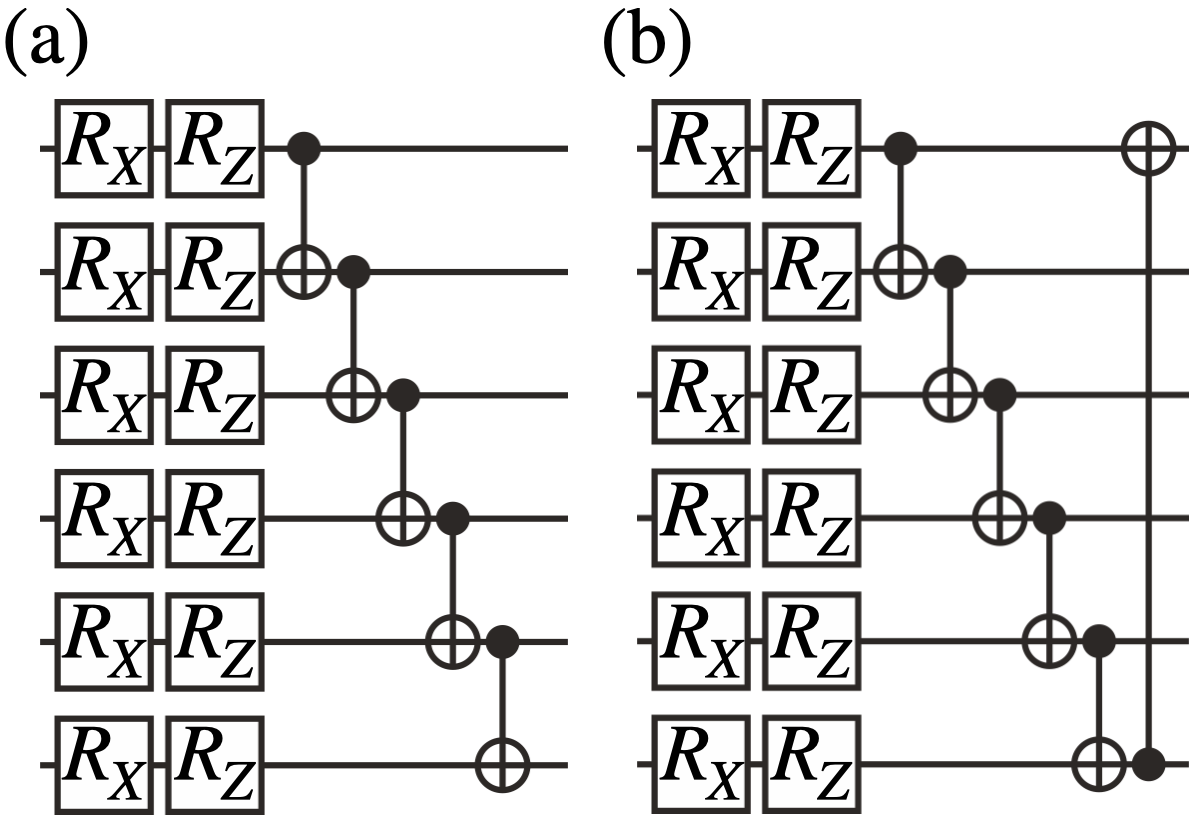}
        \caption{Two PQCs are applied as an ansatz for QNN measurements. (a) A hardware-efficient ansatz (HEA) consists of single-qubit gates and nearest-neighbor CNOT gates aligned with hardware connectivity. (b) A circuit-block (CB) unit is introduced by appending a CNOT gate at the end of a HEA, thereby increasing qubit connectivity and, consequently, enhancing the expressibility of the ansatz.}
        \label{fig:HEA}
\end{figure}

In this section, we attempt to circumvent the computational hardness of constructing a measurement circuit using a classical optimizer. We introduce QNN circuits as an ansatz for realizing quantum measurements. We, in particular, consider two types of PQCs~\cite{sim2019expressibility} in QNN measurement circuits. One is a hardware-efficient ansatz (HEA), and the other is a circuit block (CB), as shown in Fig.~\ref{fig:HEA}. An HEA is composed of two single-qubit rotations followed by nearest-neighbor CNOT gates. The structure leads to shallow quantum circuits and aligns with the typical connectivity of superconducting qubits, thereby reducing the likelihood of errors during execution. A CB unit is introduced by adding a CNOT gate at the end of an HEA and contains richer connectivity. The additional CNOT may cause a long-range interaction connecting the most distant qubits; see Fig.~\ref{fig:HEA}. The CB exhibits higher expressibility than HEA~\cite{sim2019expressibility}.

\subsection{ Hybrid Naimark-quantum-neural-network measurement} 

So far, as it is shown in Fig. \ref{fig:POVMmulti}, a Naimark measurement on $n$-qubits has two structures. One is the recipe of including ancillary qubits to construct $l$-outcome measurements. The other is an $n$-qubit measurement circuit which is described generally in $\textbf{U}(2^n)$. As mentioned in previous sections, both structures contain $ZZ$ interactions, which prevent a classical optimizer from being efficient. Here, we address this computational difficulty by relaxing one of the circuit structures, in particular, the latter one, and define and call hybrid Naimark-QNN quantum measurements as follows. 

\bigskip
{\it Hybrid Naimark-QNN measurement}: Let us consider a Naimark measurement on $n$ system qubits where a measurement circuit for $n$ qubits is described in $\textbf{U}(2^n)$. A hybrid Naimark-QNN measurement is the Naimark measurement where $n$-qubit circuits in $\textbf{U}(2^n)$ are replaced with QNN circuits. In Fig.~\ref{fig:Yordan multi qubit PQC}, a hybrid Naimark-QNN measurement for a two-qubit gate is shown, where a two-qubit gate $\textbf{U}(4)$, as shown in Fig.~\ref{fig:U(4)}, is replaced by PQC blocks.
\bigskip

\subsection{ Quantum-neural-network Measurement }

We introduce QNN measurement circuits by replacing all structures involving $ZZ$ interactions with PQCs throughout. Note that while a measurement circuit does not contain $ZZ$-interactions, the number of CNOT gates increases linearly, for which a classical optimizer may work efficiently. The PQC construction is also more favorable for near-term hardware because it has fewer CNOT gates. Since PQCs may not form a set of universal gates, multiple PQC layers are used to increase expressibility.  
 
Note that QNN quantum measurements do not explicitly control the number of POVM elements. Let us consider a QNN measurement for $l$ outcomes, and suppose that $m$ ancillary qubits are exploited. Then, overall, $2^m$ outcomes may be found, whereas $l\leq2^m$ POVM elements are needed for $l$ outcomes. 

\bigskip

 {\it QNN measurement}: Let us consider a Naimark measurement on $n$ system qubits where a measurement circuit for $n$ qubits is described in $\textbf{U}(2^n)$. After replacing $n$-qubit circuits in $\textbf{U}(2^n)$ with PQCs, we relax the remaining structure of the Naimark extension with PQCs. We call the resulting measurement circuit a fully QNN measurement, or {\it a QNN measurement} for short, which consequently contains no $ZZ$ interactions. The number of CNOT gates in a QNN measurement increases linearly. 
\bigskip

\begin{figure}[t]
        \centering
        \includegraphics[width=0.419 \textwidth]{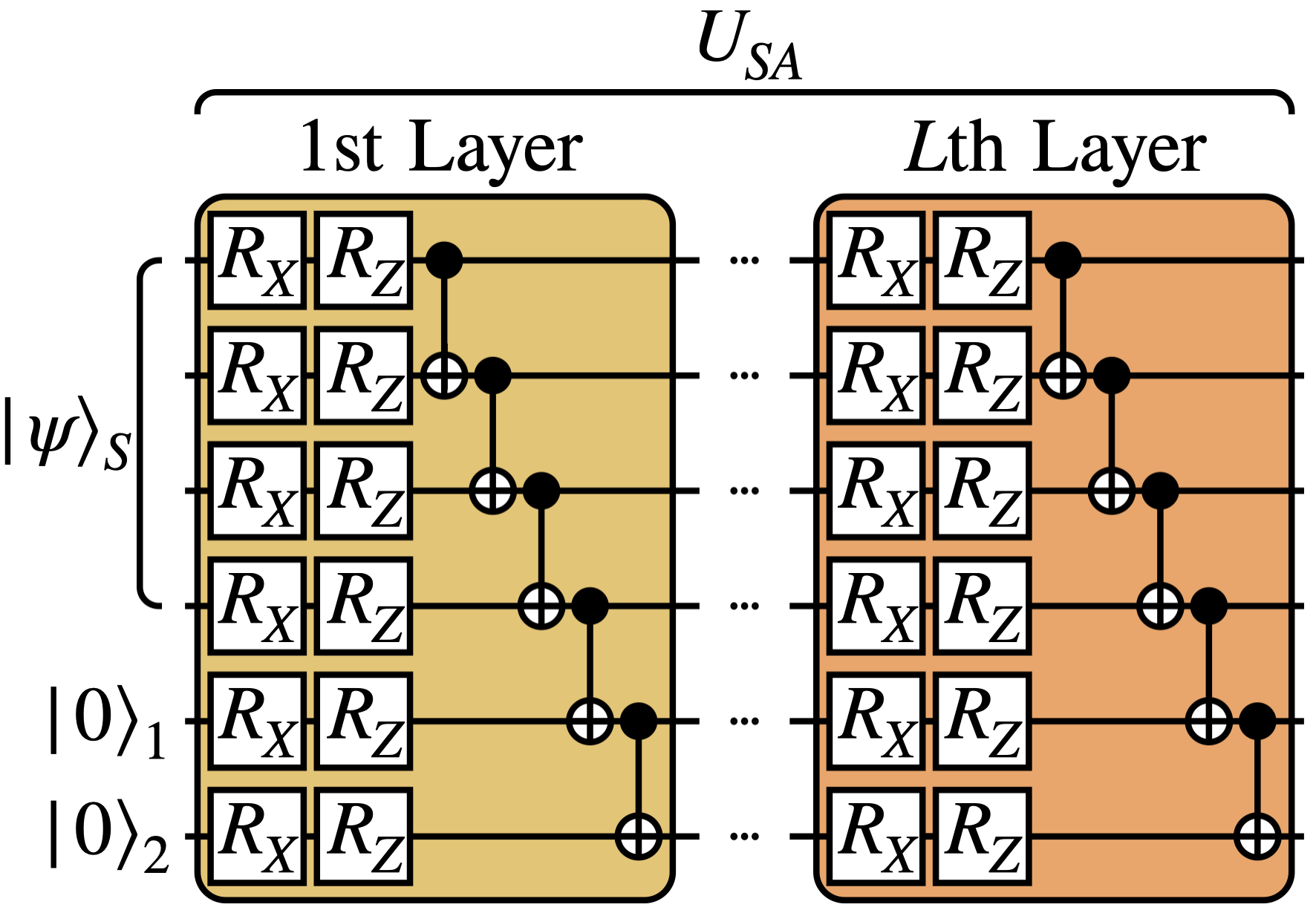}
        \caption{ A fully QNN quantum measurement circuit $U_{SA}$ is shown, where each layer is constructed by HEA. Parameters of single-qubit gates in HEA are optimized. Multiple layers may enhance QNN measurements by increasing expressibility, approximating ideal measurements.} 
     \label{fig:Neural network measurement}
\end{figure}

In Fig.~\ref{fig:Neural network measurement}, a QNN measurement for a four-qubit system is shown with HEA with two ancillary qubits. $L$ layers are applied to approximate an optimal measurement. Note also that a QNN measurement with CB can be constructed by replacing the HEA in Fig.~\ref{fig:Neural network measurement} with the CB in Fig.~\ref{fig:HEA} (b).

\section{Constructing measurement circuits for state discrimination}
\label{sec:qsd}

We here apply the constructions of measurement circuits to optimal quantum state discrimination, a fundamental task of quantum measurement \cite{Holevo1973, Yuen1975}. We consider strategies of ME~\cite{helstrom1969quantum, bergou2010discrimination, barnett2009quantum, bae2015quantum} and MC discrimination~\cite{MCM2006}. Note that MC discrimination is more general than unambiguous discrimination~\cite{barnett2009quantum} since it coincides with unambiguous state discrimination~\cite{dieks1988overlap} for linearly independent states~\cite{wm73-g73w}. 

As for the proof-of-principle demonstrations, we consider three sets of states. Note that a measurement for optimal ME discrimination is unique for an ensemble of $k$ states on $d$-dimensional Hilbert space, where $k\leq d^2$ \cite{1055941}, see also \cite{ bae2013structure}. Hence, measurements for the examples below are unique.

The first instance is a set of three states given with equal {\it a priori} probabilities, 
\bea
\mathrm{Ensemble ~1:}&& S_1 = \{|\psi_i\rangle \}_{i=0}^2~~\mathrm{where} \label{eq:en1} \\
&& |\psi_0\rangle = |0\rangle, ~~\ |\psi_1\rangle = \sqrt{\frac{2}{3}}|0\rangle+\sqrt{\frac{1}{3}}|1\rangle, \nonumber \\
&&~\mathrm{and}~~|\psi_2\rangle = \sqrt{\frac{2}{3}}|0\rangle-\sqrt{\frac{1}{3}}|1\rangle.\nonumber
\eea
Note that the analytic solution has been shown in Refs. \cite{Ban1997optimum, PeresWootters1991, Weir2018, Barnett2001, HerzogBergou2002, Bagan2012, BaeHwang2013}. The other instances are with multiple-qubit states, copies of trine states~\cite{wootters2006distinguishing, Chefles2001},
\bea
\mathrm{Ensemble ~2:} && S_2 = \{ |\psi_i\rangle =|a_i\rangle^{\otimes 2}\}_{i=0}^2 \label{eq:en2}\\
\mathrm{Ensemble ~3:} && S_3 = \{ |\psi_i\rangle =|a_i\rangle^{\otimes 3}\}_{i=0}^2 ~~\mathrm{where} \label{eq:en3} \\
&& |a_0\rangle = |0\rangle, ~~\ |a_1\rangle =  \frac{1}{2}|0\rangle + \frac{\sqrt{3}}{2}|1\rangle, \nonumber\\
&&\mathrm{and}~~ |a_2\rangle = \frac{1}{2}|0\rangle - \frac{\sqrt{3}}{2}|1\rangle. \nonumber
\eea

Although the optimal measurements for the ME-discrimination examples considered here can be realized as PVMs \cite{barnett2009quantum, bergou2010discrimination, bae2015quantum}, these examples provide useful benchmarks for evaluating how closely the proposed general POVM framework approaches the corresponding optimal discrimination performance.

In a quantum circuit, an ensemble above can be prepared with a circuit $U_{\mathrm{PREP}}$ such that 
\bea
U_{\mathrm{PREP}} |0\rangle_{P}^{\otimes \lceil\log_2 n \rceil} |0\rangle^{\otimes n}_{S} =\sum_{k=0}^{l-1} \sqrt{q_k}  |k\rangle_{P} |\psi_k\rangle_S \label{eq:prep}
\eea
and where $q_k$ is {\it a priori} probability for a state $|\psi_k\rangle$. Measuring the ancillary qubits of register $P$ in the computational basis yields the outcome $k$, thereby heralding the preparation of the corresponding state $|\psi_k\rangle$. In what follows, we construct measurement circuits with a classical optimizer. 


\subsection{Measurement for minimizing average errors}
In minimum-error (ME) discrimination, a measurement attempts to provide a guess, with minimal average error, about which state has been prepared among an ensemble $S = \{q_i, \rho_i \}$, which describes a preparation of a state $\rho_i$ with {\it a priori} probability $q_i$. The guessing probability is given by,
\bea
p_{\mathrm{guess}}^{(S)} = \max_{\{M_i\}} \sum_i q_i \tr[M_i \rho_i].  \label{eq:med}
\eea
where $\{M_i\}$ forms a complete measurement. For the ensembles above, the guessing probability can be computed analytically \cite{Andersson2002mimunum, Barnett2001, Weir2018, bae2013structure, Ban1997}: 
\bea
p_{\mathrm{guess}}^{(S_1)} = \frac{2}{3},~ 
p_{\mathrm{guess}}^{(S_2)} = \frac{3+2\sqrt{2}}{6} \approx 0.9714,~\mathrm{and}~ \nonumber\\p_{\mathrm{guess}}^{(S_3)} = \frac{ 7+2\sqrt{6}}{12} \approx 0.9916. \label{eq:analytic} 
\eea
More copies lead to higher distinguishability.

\begin{algorithm}[h]
\caption{Variational optimization of minimum-error discrimination}
\label{alg:MED}
\begin{algorithmic}[1]

\Require Ensemble $\{(q_i,|\psi\rangle_i)\}_{i=0}^{l-1}$
\Require Measurement Ansatz $U(\boldsymbol{\theta})$
\Require Initial parameters $\boldsymbol{\theta}_0$
\Require Classical optimizer
\Require Objective function $\mathcal{L}(\boldsymbol{\theta})=1/P_{\mathrm{guess}}(\boldsymbol{\theta})$

\Ensure Optimized parameters $\boldsymbol{\theta}^{*}$
\Ensure Optimized guessing probability $P_{\mathrm{guess}}(\boldsymbol{\theta}^{*})$

\State Initialize $\boldsymbol{\theta} \gets \boldsymbol{\theta}_0$

\While{convergence criterion is not satisfied}

    \State Prepare the input state $\sum_{k=0}^{l-1} \sqrt{q_k}  |k\rangle_{P}  |\psi_k\rangle_S$
    \State Initialize $\lceil\log_2 l\rceil$ ancillary qubits in $\ket{0}_A^{\otimes \lceil\log_2 l\rceil}$
    \State Apply $U(\boldsymbol{\theta})$
    \State Measure the ancillary qubits of register $P$ and $A$ and estimate $P_{\mathrm{guess}}$
    \State Evaluate $\mathcal{L}(\boldsymbol{\theta})=1/P_{\mathrm{guess}}(\boldsymbol{\theta})$
    \State Update $\boldsymbol{\theta}$ using the classical optimizer

\EndWhile

\State $\boldsymbol{\theta}^{*} \gets \boldsymbol{\theta}$

\State \Return $\boldsymbol{\theta}^{*}$ and
$P_{\mathrm{guess}}(\boldsymbol{\theta}^{*})$

\end{algorithmic}
\end{algorithm}

In what follows, we construct optimal measurements with {\it measurement ansatzes} as follows. 
\begin{itemize}
    \item {\it Naimark measurements with a classical optimizer}. Firstly, we exploit the structure of a Naimark measurement as shown in Figs.~\ref{fig: POVMsingle} and~\ref{fig:POVMmulti}. In the construction, we leave single-qubit gates with parameters to be fixed, which are then trained on the circuit with a classical optimizer to find optimal parameters. Once a set of parameters is fixed, we have a measurement for ME discrimination. 
    
    \item {\it hybrid Naimark-QNN measurements with a classical optimizer}. Secondly, we keep the structure of Naimark measurements and apply QNN circuits on system qubits, which we call hybrid Naimark-QNN measurements. For the ensemble $S_2$ in Eq.~(\ref{eq:en2}), we replace two-qubit exact gates in $\text{U}(4)$ with PQCs and apply a classical optimizer. 
    
    \item {\it Fully QNN measurements with a classical optimizer}. Thirdly, we attempt to construct POVMs with QNN measurements in Fig.~\ref{fig:Neural network measurement} for ensembles $S_1$, $S_2$, and $S_3$ by using a classical optimizer. We consider HEA-based QNN measurements (QNN-HEA) and CB-based QNN measurements (QNN-CB).
\end{itemize}

Algorithm \ref{alg:MED} summarizes the overall optimization procedure for ME discrimination. In the algorithm, the ansatz $U(\boldsymbol{\theta})$ corresponds to one of the three approaches described above.

\subsubsection*{Resources: } 
CNOT gates in QNN measurements increase linearly in the number of qubits. The number of CNOT gates in a Naimark measurement circuit is given as $O(2^n l^2)$. Implementation of $l$ outcomes requires $n_A = \lceil \log_2 l\rceil$ ancillary qubits, where $l\in ( 2^{n_A-1}, 2^{n_A} ]$. The count is derived in detail in the Appendix \ref{sec:appcnot}. When $n_A$ is dominant, i.e., a larger number of outcomes is needed, the overall number of CNOT gates increases rapidly. 

The number of CNOT gates can be compared with the other construction in Ref.~\cite{DKPpaper}, where the number is given by $O(2^{2n}l)$ with $l = 2^{n_A}$. The count increases rapidly as a system contains more qubits; however, the overall count is less dependent on ancillary qubits. 

It is worth mentioning that ME discrimination of $n$-qubit states does not have to contain more than $(2^n)^2$ outcomes to achieve the guessing probability in general, even if the number of states $l$ is larger than $(2^{n})^2$~\cite{1055941}. For instance, one does not need to have more than $4$ outcomes in ME discrimination of $l$ qubit states for arbitrary $l\geq 2$. Hence, one can restrict ${n_A}\leq {2n} $ without loss of generality.

\subsubsection*{ Proof-of-principle demonstration:}
\begin{figure}[h]
        \centering
     \includegraphics[width=0.5\textwidth]{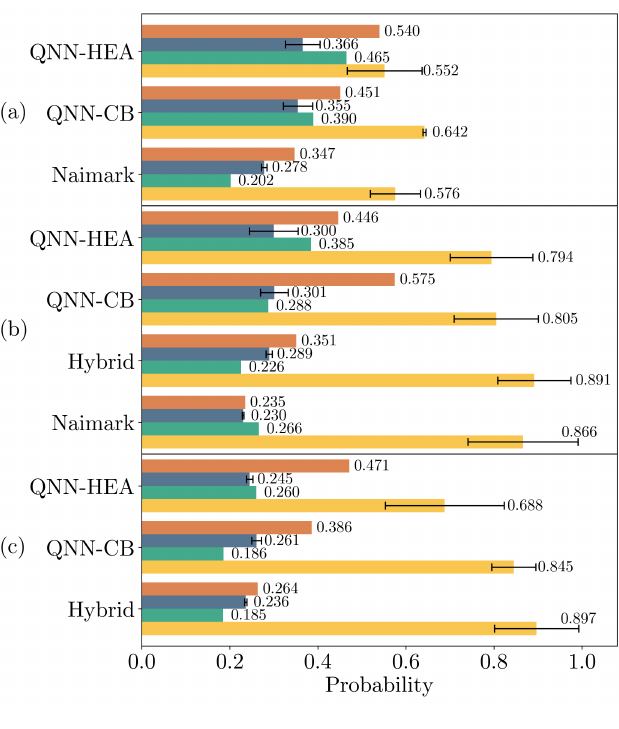}
        \caption{ Measurement ansatzes, such as Naimark, hybrid Naimark-QNN, and QNN measurement circuits (with layers $L=3$), are optimized and applied for state discrimination for ensembles $S_1$ in (a), $S_2$ in (b), and $S_3$ in (c). They are optimized in i) (yellow) the simulator Qiskit Aer, ii) (blue) IQM Garnet on 28/08/2026, and iii) (green) the IBMQ ibm$\_$strasbourg on 29/11/2024. A mixed strategy is to realize a measurement ansatz with optimal parameters obtained from the simulator (red). } 
        \label{fig:overall}
\end{figure}

We realize the construction of optimal measurements, via the measurement ansatzes, in i) the Qiskit Aer Simulator, ii) the IQM Garnet, and iii) the IBMQ ibm$\_$strasbourg. iv) A mixed strategy is to realize a measurement ansatz with optimal parameters obtained from the simulator. For QNN-HEA and QNN-CB, we attempt $3$ layers. The results are shown and summarized in Fig.~\ref{fig:overall}. For each ansatz, we have $30$ runs with different seeded initial parameters. Unless otherwise specified, COBYLA is applied as a classical optimizer with a convergence tolerance of $10^{-8}$, and the objective function is estimated using $10^4$ shots. One may consider various optimizers, other than COBYLA, such as BFGS, SLSQP, and SPSA. No particular dependence or bias on an optimizer is observed; the comparison is shown in Table \ref{tab:optimizer_results}.

\begin{table}[h]
\centering
\small
\caption{Maximum values of guessing probabilities are collected by varying classical optimizers when constructing optimal measurements for ensemble $S_2$ via the hybrid Naimark-QNN measurements. No particular bias is observed. Note that $\delta p_{\mathrm{guess}}$ denotes a standard deviation. }
\label{tab:optimizer_results}
\renewcommand{\arraystretch}{1.2}
\setlength{\tabcolsep}{3pt}
\begin{tabular}{c|cccccc}
\hline
\textbf{Optimizer~~}
& \textbf{COBYLA~~}
& \textbf{BFGS~~}
& \textbf{Powell~~}
& \textbf{SLSQP~~}
& \textbf{SPSA}
\\
\hline
$p_{\mathrm{guess}}^{(S_2)}$
& 0.951
& 0.941
& 0.951
& 0.948
& 0.952
\\
\hline
\begin{tabular}[c]{@{}c@{}}
 $\delta p_{\mathrm{guess}}^{(S_2)} $
\end{tabular}
& 0.083 
& 0.164 
& 0.079
& 0.189 
& 0.084 
\\
\hline
\end{tabular}
\end{table}

Let us summarize the observations from Fig. \ref{fig:overall}. First, optimizing parameters in a Naimark ansatz and a hybrid Naimark-QNN ansatz in a simulator finds measurements that give the guessing probability in Eq. (\ref{eq:analytic}). Optimizing parameters in QNN measurements, both HEA and CB cases, does not achieve the guessing probability in Eq.(\ref{eq:analytic}). Second, in the mixed-strategy case, implementations with a Naimark measurement show the smallest guessing probability among all instances. This is because a Naimark measurement has a larger circuit depth than the others. Hence, the low guessing probabilities are due to hardware noise. Third, when training parameters directly on real quantum hardware, QNN measurements show higher guessing probabilities than the others. QNN-HEA shows a higher guessing probability than QNN-CB. We also summarize ME discrimination on IQM Garnet in Table~\ref{tab:ME_IQM}.

\begin{table*}[t]
\centering
\small
\caption{ Proof-of-principle demonstrations of ME discrimination for ensembles $S_1$, $S_2$, and $S_3$ on IQM Garnet are summarized: parameters, single- and two-qubit gates are counted, and transpiled circuit depth is shown. Standard deviations of guessing probabilities are denoted by $\delta p_{\mathrm{guess}}$. }
\label{tab:ME_IQM}
\renewcommand{\arraystretch}{1.1}
\setlength{\tabcolsep}{2pt}
\begin{tabular}{c|ccc|cccc|ccc}
\hline
\textbf{Ensemble}
& \multicolumn{3}{c|}{\textbf{$S_1$}}
& \multicolumn{4}{c|}{\textbf{$S_2$}}
& \multicolumn{3}{c}{\textbf{$S_3$}}
\\
\hline
\textbf{Ansatz} &
\textbf{HEA} &
\textbf{CB} &
\textbf{Naimark} &
\textbf{HEA} &
\textbf{CB} &
\textbf{Hybrid} &
\textbf{Naimark} &
\textbf{HEA} &
\textbf{CB} &
\textbf{Hybrid}
\\
\hline

\begin{tabular}[c]{@{}c@{}}
Trainable\\
parameters ($\#$)
\end{tabular}
&18 &18 &19
&24 &24 &28 &83
&30 &30 &46 \\
\hline

\begin{tabular}[c]{@{}c@{}}
Single-qubit \\
gates ($\#$)
\end{tabular}
&55 &92 &298
&94 &102 &614 &1754
&156 &180 &1173 \\
\hline

\begin{tabular}[c]{@{}c@{}}
Two-qubit \\
gates ($\#$)
\end{tabular}
&11 &24 &91
&24 &27 &207 &558
&44 &58 &390 \\
\hline

Circuit depth
&41 &73 &251
&75 &85 &488 &1483
&138 &166 &938 \\
\hline


\begin{tabular}[c]{@{}c@{}}
$p_{\mathrm{guess}}$
\end{tabular}
&0.366 &0.355 &0.278 
&0.300 &0.301  &0.290 &0.230
&0.245 &0.261 &0.236 \\
\hline

\begin{tabular}[c]{@{}c@{}}
$\delta p_{\mathrm{guess}}$
\end{tabular}
&0.039  &0.033  &0.006 
&0.055  &0.031  &0.007  &0.002
&0.007  &0.011 &0.003 \\
\hline

\end{tabular}
\end{table*}

\begin{figure*}[t]
    \centering
    \includegraphics[width=0.95 \textwidth]{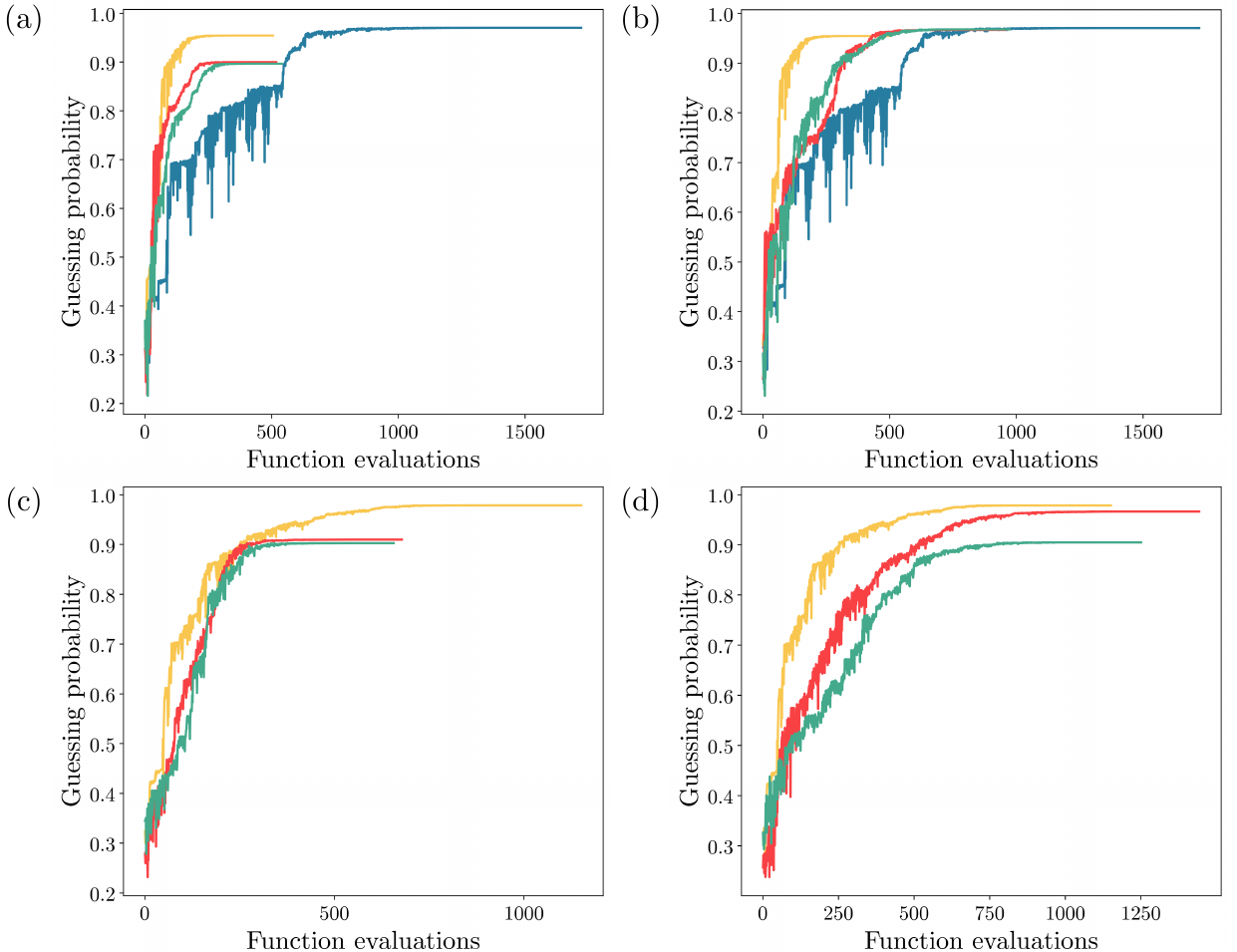} 
    \caption{ Measurements for ME discrimination for ensembles $S_2$ and $S_3$ are constructed by optimizing the parameters of the measurement ansatzes, such as Naimark (blue), hybrid Naimark-QNN (yellow), QNN-HEA (red) and QNN-CB (green) measurements. QNN measurements may rely on the number of layers $L$. Measurement ansatzes are trained for ME discrimination (a) for ensemble $S_2$, where QNN measurements exploit three layers ($L=3$), (b) for ensemble $S_2$, with QNN measurements having five layers ($L=5$), (c) for ensemble $S_3$ with three layers ($L=3$), and (d) for ensemble $S_3$ with five layers ($L=5$). Note that a Naimark measurement is not constructed for ensemble $S_3$ due to the computational hardness of the optimization.}
    \label{fig:convergence}
\end{figure*}

We also observe that a classical optimizer eventually constructs an optimal measurement with a Naimark ansatz. However, QNN measurement circuits may not reach the guessing probability of an optimal measurement. One may attempt to improve QNN measurements by using more layers. For a few-qubit task, $5$ layers often achieve convergence \cite{sim2019expressibility}.

\begin{table*}[t]
\centering
\small
\caption{Results and the transpiled circuit depth for ME discrimination for ensembles $S_1$, $S_2$, and $S_3$ on ibm$\_$strasbourg.}
\label{tab:ME_IBM}
\renewcommand{\arraystretch}{1}
\setlength{\tabcolsep}{4pt}
\begin{tabular}{c|ccc|cccc|ccc}
\hline
\textbf{Ensemble}
& \multicolumn{3}{c|}{\textbf{$S_1$}}
& \multicolumn{4}{c|}{\textbf{$S_2$}}
& \multicolumn{3}{c}{\textbf{$S_3$}}
\\
\hline
\textbf{Ansatz}
& \textbf{HEA}
& \textbf{CB}
& \textbf{Naimark}
& \textbf{HEA}
& \textbf{CB}
& \textbf{Hybrid}
& \textbf{Naimark}
& \textbf{HEA}
& \textbf{CB}
& \textbf{Hybrid}
\\
\hline
Circuit depth
& 65
& 75
& 724
& 163
& 232
& 1449
& 4867
& 238
& 384
& 2431
\\
\hline
$p_{\mathrm{guess}}$
&0.465
&0.390
&0.202
&0.385
&0.288
&0.226
&0.266
&0.260
&0.186
&0.185
\\
\hline

\end{tabular}
\end{table*}

In Fig. \ref{fig:convergence}, the results comparing $3$- and $5$-layer QNN ansatzes are shown and summarized. To observe the convergence behavior, we run the optimization in the simulator. First, QNN measurements with a $3$-layer ansatz for ensemble $S_2$ do not reach the convergence of a Naimark ansatz; see Fig. \ref{fig:convergence} (a). Second, QNN measurements are improved by more layers: with $5$-layer ansatz, the QNN ansatz is very close to the convergence of a Naimark measurement; see Fig. \ref{fig:convergence} (b), where the gap with a Naimark ansatz remains $0.002$; the gap is detected after running $20$ instances of QNN-HEA and Naimark measurements, where the former obtains a guess probability of $0.963$ and the latter $0.965$. For ensemble $S_3$, see Fig. \ref{fig:convergence} (c) and (d); the optimization of parameters with a Naimark ansatz is not feasible, but Hybrid Naimark-QNN and QNN measurements are. QNN-HEA and QNN-CB with $3$ layers may show similar convergence behavior. When $5$ layers are applied, QNN-HEA achieves a measurement showing a higher guessing probability than QNN-CB.

From the proof-of-principle demonstrations shown in Figs. \ref{fig:overall} and \ref{fig:convergence}, QNN measurement circuits with HEA are the most favorable for an optimizer to work with real and noisy quantum hardware. The second most favorable one is also a QNN measurement with CB. The next one is a hybrid Naimark-QNN measurement, which keeps the structure of the Naimark extension.

We observe that the key distinction lies in the distribution of CNOT gates in a circuit. Both a hybrid Naimark-QNN measurement and a Naimark measurement share an identical arrangement of CNOT gates, where CNOT gates are not sequential between qubits: CNOT operations over qubits far apart are also required. Fully QNN measurements with HEA contain CNOT gates between nearest qubits; CB is constructed such that after HEA it includes a single CNOT for two qubits that are not nearest neighbors.

\subsection{Maximum confidence measurement}

Let us consider the other strategy of discriminating quantum states, a maximum-confidence (MC) measurement, that generalizes strategies for state discrimination~\cite{MCM2006, barnett2009quantum}. For an ensemble $\{q_{i},\rho_i \}_{i}^m$, i.e., a state $\rho_i$ appears with a probability $q_i$. Suppose that an outcome $k$ occurs, which makes a guess that a state $\rho_k$ has been generated. An MC measurement introduces the notion of confidence per a detection event. The confidence, denoted by $C(k)$, is defined as a conditional probability as follows \cite{MCM2006}, 
\bea
C(k)= p_{P|M} (k| k) = \frac{ p_{P}( k) ~p_{M|P}( k |k)}{p_M(k)} \nonumber
\eea
where $P$ and $M$ denote preparation and a measurement, respectively. Since we have $p_P (k) = q_k$, $p_{M|P} (k|k) = \tr[\rho_k M_k]$ and $p_{M } ( k) = \tr[\rho M_k] $, where $\rho = \sum_i q_i \rho_i$, the confidence above maximized over all POVM elements is given by \cite{MCM2006}
\bea
\max C(k) = \max_{M_k} \frac{q_k \mathrm{tr}[\rho_k M_k] }{\mathrm{tr}[\rho M_k]}
\label{eq: MCM}
\eea
which asks to optimize a POVM element $M_k$.

Compared to ME discrimination that maximizes an average guessing probability, an MC measurement maximizes the probability that a guess is correct {\it per} outcome \cite{MCM2006}. For instance, maximizing the confidence over all states corresponds to ME discrimination, with the constraint that all outcomes are conclusive \cite{barnett2009quantum}.

The optimization problem has been analytically solved for qubit states~\cite{MCMlee} and is generally approached via a semidefinite program~\cite{flatt2022contextual}. 

The optimization problem can be facilitated by taking an ansatz $M_k = c_k \rho^{-\frac{1}{2}} Q_k \rho^{-\frac{1}{2}}$ for some $Q_i\geq 0$ and constant $c_i$ so that the problem becomes linear 
\bea
C(k)= \max_{Q_k \geq0,~ \mathrm{tr}[Q_k]=1} \mathrm{tr}[ \rho^{-\frac{1}{2}} q_k \rho_k \rho^{-\frac{1}{2}} Q_k]. \label{eq: MCM linear}
\eea
The positive constants $\{c_i\}$ determine the weights of the conclusive outcomes such that $\sum_i M_i\leq I$. The inconclusive outcome is given by 
\bea
M_? = I - \sum_i M_i. 
\eea
The probability of the inconclusive outcome is calculated as 
\bea
p_? = \mathrm{tr}[\rho M_?] = 1 - \sum_i \mathrm{tr}[\rho M_i]. \label{eq:pin}
\eea
For the maximization of confidence in Eq. (\ref{eq: MCM linear}), it suffices to consider a rank-one projector for $Q_k$, which simplifies the problem to finding a unitary circuit $U$ such that $Q_k={U}\ketbra{k}{U}^{\dagger}$. After all, we can formulate MCMs for an ensemble $\{q_i,\rho_i \}$ as follows, 
\bea 
\max_{U} \mathrm{tr}[{U}^\dagger \rho^{-\frac{1}{2}} q_k \rho_k \rho^{-\frac{1}{2}} {U} |k \rangle \langle k|]. \label{eq: MCM final}
\eea
Therefore, we arrive at the problem of constructing a unitary circuit $U$ for an ensemble.

The optimization problem in Eq.~(\ref{eq: MCM final}) contains two steps: firstly, we realize a projection onto the subspace spanned by $\sqrt{\rho}^{-1}$, and secondly, after the projection is successful, we optimize a unitary circuit $U$. The former corresponds to state filtration by a Kraus operator proportional to $\sqrt{\rho}^{-1}$ and can be realized by block encoding (BE), and the latter by a classical optimizer. 

We write by $U_{BE}$ for the purpose, as follows, 
\begin{align}
&U_{BE} \left[ \sum_{k=0}^{l-1}\sqrt{q_k}|k\rangle_{A_1} |\psi_k\rangle_S \otimes |0\rangle^{\otimes n+1}_{A_2} \right] \nonumber\\
&= \sum_{k=0}^{l-1} \sqrt{q_k}|k\rangle_{A_1} K |\psi_k\rangle_S \otimes|0\rangle^{\otimes n+1}_{A_2} + |\Phi^\bot\rangle \label{eq:bec}
\end{align}
where $A_1$ denote ancillary qubits for preparing an ensemble $\{q_k, |\psi_k\rangle \langle \psi_k| \}$, $|\Phi^{\perp}\rangle$ is given such that $_{A_2}\langle 0|^{\otimes n+1}|\Phi^{\perp}\rangle =0$, and 
\bea
K = \frac{1}{c } \rho^{-1/2} ~\mathrm{where}~ c= 2^n \max_{i,j} | (\rho^{-1/2})_{ij} |. 
\nonumber 
\eea
Note that $K^{\dagger}K\leq I$ for the construction above since we have $\|K \|_{2}^{2} \leq \|K \|_1 \|K \|_{\infty}$ in which $\|K \|_1, \|K \|_{\infty} \leq 1$ given $K=[K_{ij}]$ where $|K_{ij}|\leq 1/2^n$. Measurement outcomes $0^{\times (n+1)}$ in ancillary qubits of the register $A_2$ realizes a transformation $\rho^{-1/2}$ on system qubits, see Eq.~(\ref{eq: MCM final}). In the following, we summarize strategies for block-encoding and unitary optimization, see Fig. (\ref{fig: MCM BE}).

 \begin{figure}[t]
    \centering
    \includegraphics[width=0.45\textwidth]{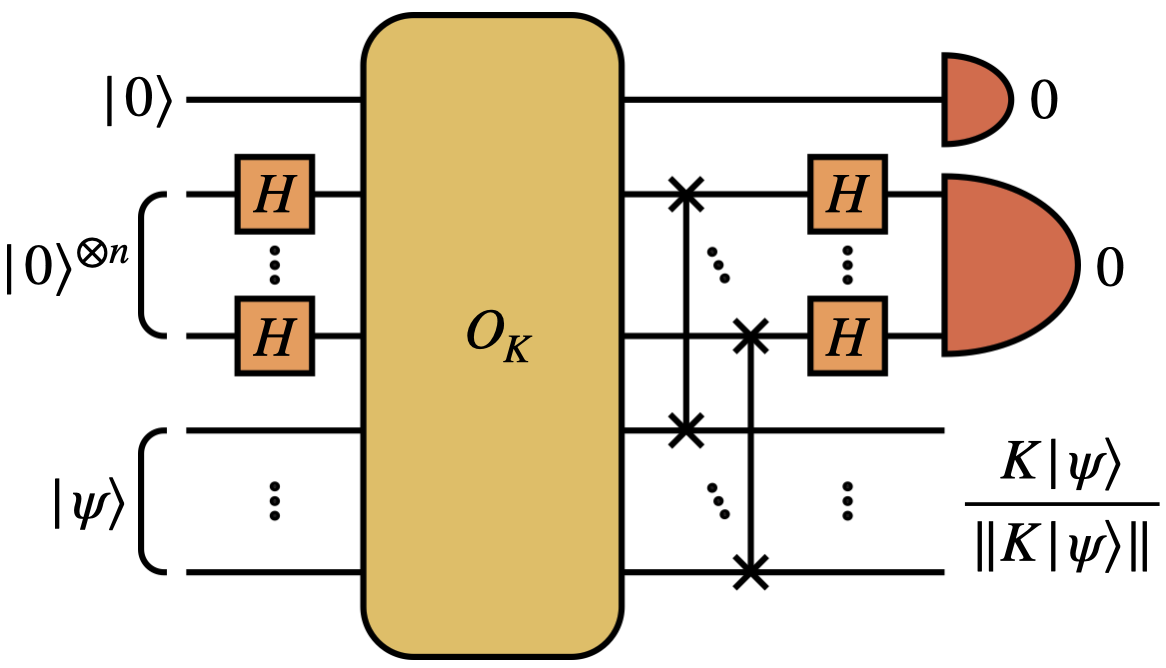}
        \caption{ The block-encoding method is shown for the realization of Eq.~(\ref{eq:bec}). A Kraus operator $K$ is realized on system qubits when measurement outcomes in the ancillary qubits are all zeros.} \label{fig:bee}
\end{figure}

\subsubsection*{ Block-encoding }
    
 \begin{figure}[t]
    \centering
    \includegraphics[width=0.47\textwidth]{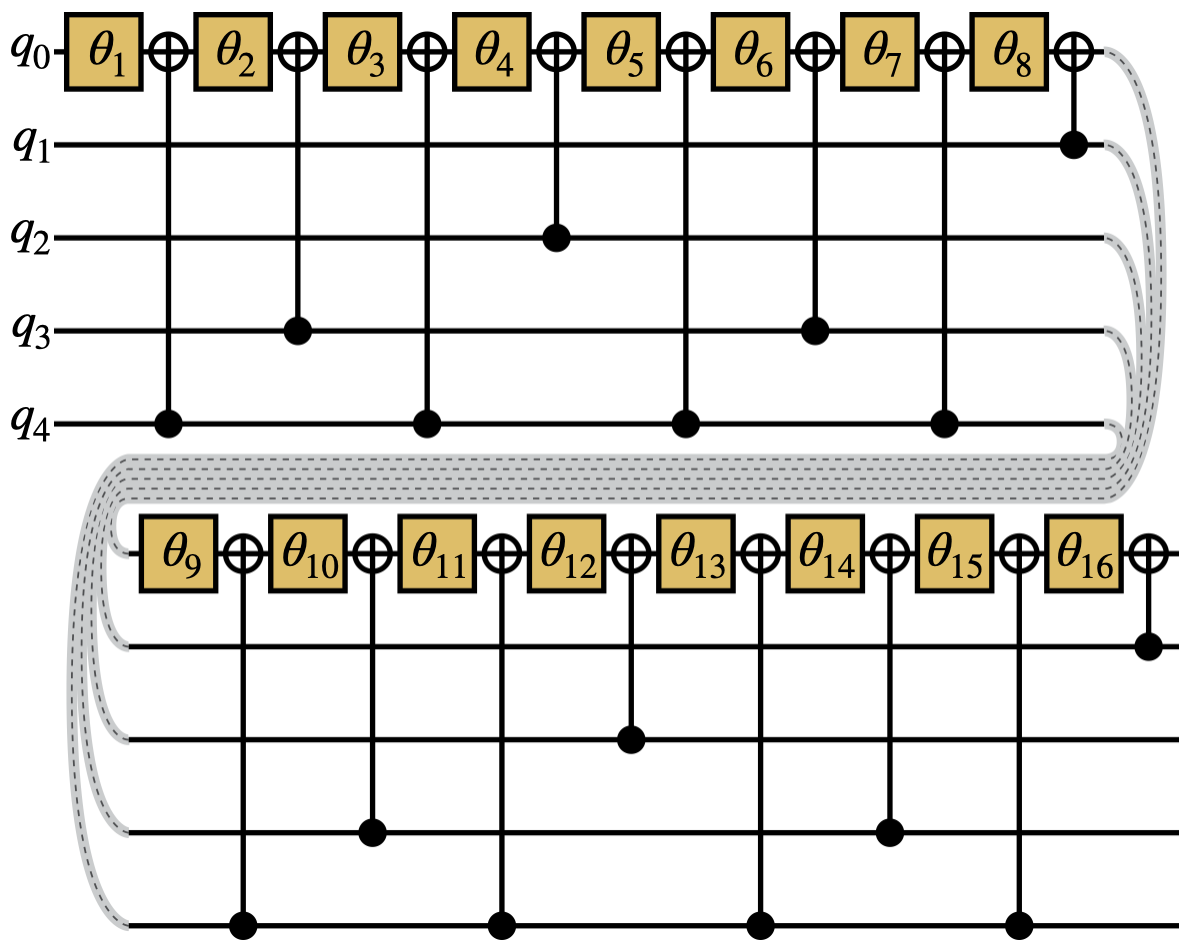}
        \caption{ A unitary transformation $O_K$ in the block-encoding in Fig.~\ref{fig:bee} is shown for system qubits $q_3$ and $q_4$ with ancillary qubits $q_0, q_1$ and $q_2$. A Gray code \cite{Sawaya2020} applies to qubits in registers $q_1q_2q_3q_4$ and CNOT gates are placed between $q_0$ and $q_i$ for $i\in \{1,2,3,4\}$. For instance, starting from the Gray code $0000$, a CNOT gate is placed between $q_0$ and $q_3$ for the code $0011$.}
        \label{fig:BEOA}
\end{figure}

To implement a block-encoding $K$, we employ the method shown in Ref.~\cite{9951292}, see Fig.~\ref{fig:bee}. Let $U_K$ denote a unitary circuit for the block-encoding that implements an $n$-qubit Kraus operator $K$ 
\bea
K = \Big( \langle0|^{\otimes n+1} \otimes I_n\Big) U_{K} \Big( |0\rangle^{\otimes n+1} \otimes I_n\Big) 
\eea
where $U_K$ contains Hadamard and SWAP gates as well as a ($2n+1$)-gate $O_K$ that works as follows 
\bea 
O_K |0\rangle |i\rangle |j\rangle =\Big(2^n K_{ij} |0\rangle + \sqrt{1- 4^n K_{ij}^2} |1\rangle \Big) |i\rangle |j\rangle \nonumber
\eea 
for $n$-qubit states $|i\rangle$ and $|j\rangle$. One can implement $O_K$ by using \(2^n \times 2^n\) controlled-\(R_Y\) gates, 
\bea
O_K = \sum_{i, \ j=1}^{2^n} R_y(2\theta_{ij}) \otimes |i\rangle \langle i| \otimes |j\rangle \langle j| \nonumber
\eea
where $\theta_{ij}=\arccos(2^n K_{ij})$. One can construct $O_K$ with CNOTs and single-qubit gates; see Fig.~\ref{fig:BEOA}. A circuit implementation of $O_K$ may introduce another parameterization, denoted by $\hat{\theta}$, where we apply a cutoff condition that small rotations below a threshold, i.e., $\hat{\theta}_{ij}\leq\delta_c$, are ignored. We write by $\widetilde{K}$ the Kraus operator resulting from the final circuit implementation, including the cutoff constraint, and the approximation can be estimated by the cutoff parameter, $\Vert K- \widetilde K\Vert_2\leq 4^n\delta_c$~\cite{9951292}.

\subsubsection*{Optimizing a unitary circuit}

After block-encoding is successfully realized, the next step is to optimize a unitary circuit for an MC measurement, i.e., $U$ in Eq.~(\ref{eq: MCM final}). A circuit for the optimization generally requires universal gates such as single-qubit rotations and CNOT gates, where we call an exact construction a Naimark measurement, in a manner similar to ME discrimination. Note that the exact construction coincides with measurements for ME discrimination when an MCM is constrained to have no inconclusive measurement outcomes. 

Let $U:= U(\Theta)$ denote a circuit with a set of parameters $\Theta$, either a universal circuit or a QNN measurement circuit. Firstly, one can consider optimizing parameters for universal gates and may ultimately achieve the highest MC values. Secondly, one can efficiently exploit a classical optimizer by relaxing a circuit with universal gates with QNN measurement circuits. Similar to cases of ME discrimination, we exploit fully QNN measurement circuits with HEA and CB.

\begin{figure}
    \centering
    \includegraphics[width=0.5\textwidth]{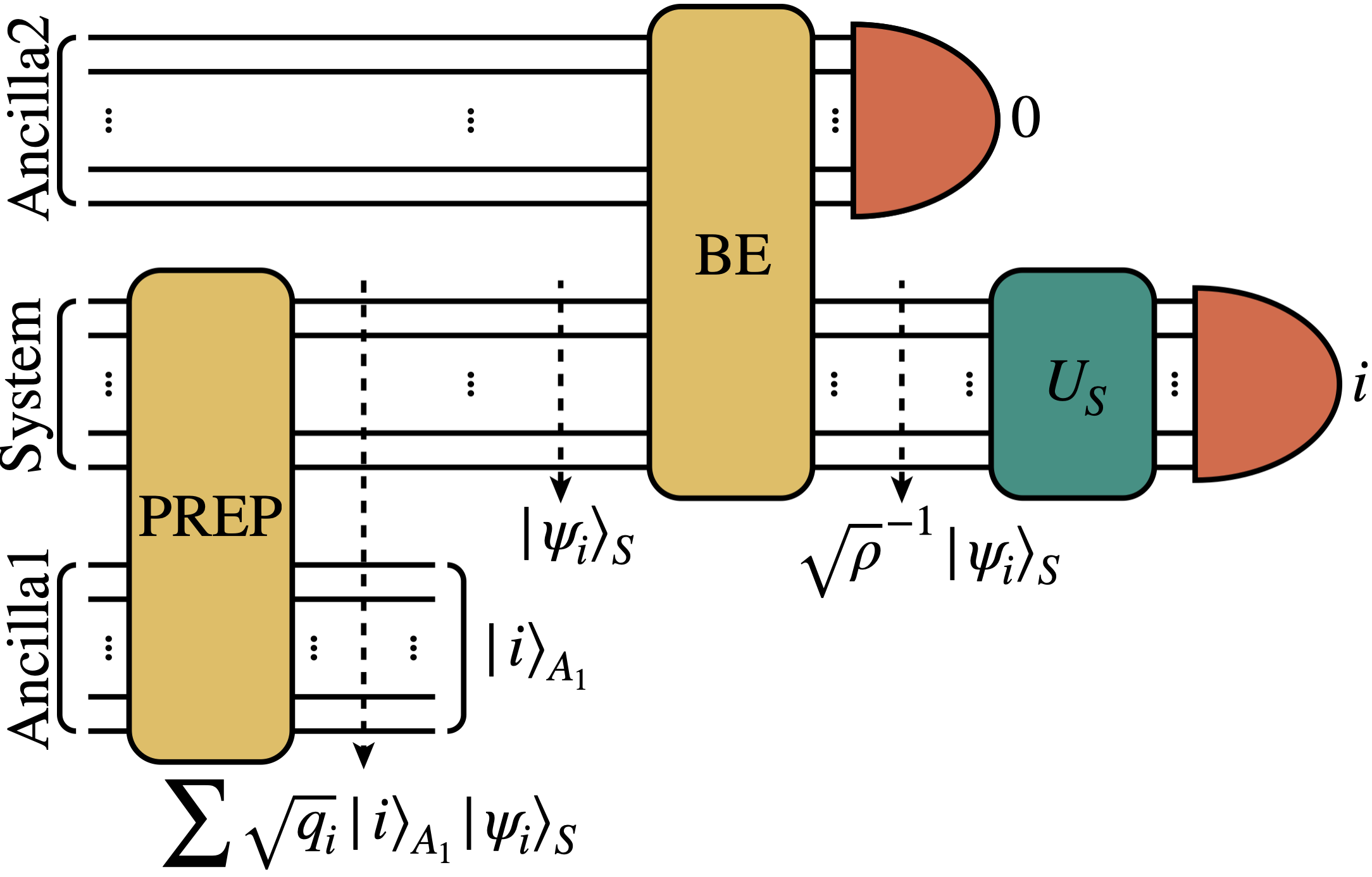}
    \caption{ The circuit of the proof-of-demonstrations of MC discrimination is summarized. An ensemble is prepared by a unitary circuit PREP; measurement $i$ occurs with {\it a priori} probability $q_i$, and prepares a state $|\psi_i\rangle $ on system qubits. A circuit BE realizes a block-encoding that implements a projection $\rho^{-1/2}$; a measurement outcome $0$ concludes a successful realization of the projection on system qubits, see Eq. (\ref{eq:bec}). Once a block encoding is applied, a unitary circuit $U_S$ is optimized to maximize the confidence values.}
    \label{fig: MCM BE}
\end{figure}

\subsubsection*{ Optimizing the inconclusive outcome: } Once we obtain a unitary operator $U(\Theta)$ for MC discrimination by optimizing measurement ansatzes, we can compute the rate of inconclusive outcomes as follows. From the set of parameters $\Theta$, we can reproduce $\rho^{-1/2}Q_i( \Theta)\rho^{-1/2}$ which is proportional to $M_i$ up to a constant. From Eq. (\ref{eq:pin}), the minimal probability of inconclusive outcomes can be found by solving the minimization problem: 
\begin{equation}
    \min_{ \{ c_{i}\geq 0\} } \| \sum_i c_i \rho^{-1/2}Q_i( \Theta) \rho^{-1/2} - I\|_1. 
\end{equation}
One can apply classical optimizers, such as Sequential Least SQuares Programming(SLSQP).

\subsubsection*{ Proof-of-principle demonstration:} We consider MC discrimination for the ensembles $S_1$, $S_2,$ and $S_3$ in Eqs.~(\ref{eq:en1}), (\ref{eq:en2}), and (\ref{eq:en3}), respectively. We optimize the parameters of the measurement ansatz in two ways: via the simulator, Qiskit Aer, and real quantum hardware, \texttt{ibm\_strasbourg}.

For each ensemble, we first implement the FABLE for block-encoding, and then optimize a unitary gate $U(\Theta)$ by Naimark or QNN measurement circuits \cite{Camps_2022}. Note that whenever $\rho$ is rank deficient, we replace its inverse with the Moore–Penrose pseudoinverse. Note that we have $C_{\max}=3/7$ for the state $|\psi_0\rangle$ in $S_1$ \cite{MCM2006, MCMlee}. The states in ensemble $S_2$ are linearly independent, so $C_{\max}=1$. For the same reason, we have $C_{\max}=1$ for states in ensemble $S_3$.

\begin{table*}[t]
\centering
\small
\caption{Results and the transpiled circuit depth of MC discrimination for ensembles $S_1$, $S_2$, and $S_3$ on ibm$\_$strasbourg.}
\label{tab:transpiled_depth_mcm}

\renewcommand{\arraystretch}{1.2}
\setlength{\tabcolsep}{1pt}

\begin{tabular}{c|cc|cc|cc}
\hline

\textbf{Ensemble}
& \multicolumn{2}{c|}{\textbf{$S_1$}}
& \multicolumn{2}{c|}{\textbf{$S_2$}}
& \multicolumn{2}{c}{\textbf{$S_3$}}
\\
\hline

\textbf{Ansatz}
& \textbf{QNN-HEA~~}
& \textbf{Naimark~~}
& \textbf{QNN-HEA~~}
& \textbf{Naimark~~}
& \textbf{QNN-HEA~~}
& \textbf{QNN-CB}
\\
\hline

$\delta_c$
&0.005
&0.005
&0.005
&0.005
&0.025
&0.025
\\
\hline
Circuit depth
& 104
& 89
& 207
n
 
& 239
& 687
& 665
\\
\hline
$C_{\max}$
&0.283
&0.331
&0.611
&0.500
&0.227
&0.174
\\
\hline
\end{tabular}
\end{table*}

For block-encoding, we ignore single-qubit gates having sufficiently small angles $\hat\theta_{ij}\leq \delta_c$ for efficient realizations. Discarding small-angle rotations introduces errors in the approximation: we take a cut-off $\delta_c =0.005$ $1$ and $2$-qubit cases to maintain the accuracy. In the $3$-qubit case, due to the significant circuit depth caused by serial application of \(2^n \times 2^n\) CNOT gates, we set $\delta_c=0.025$ for the real quantum hardware, whereas we report results for both $\delta_c=0.005$ and $\delta_c=0.025$ on the simulator.

\begin{figure}[t] 
\centering
\includegraphics[width=0.5\textwidth]{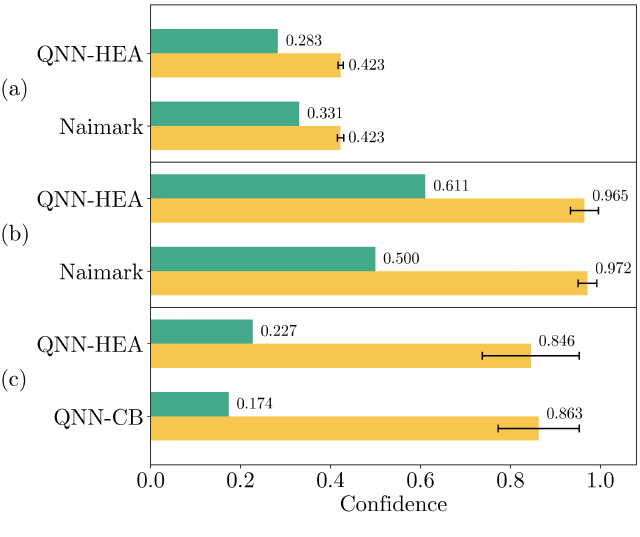}
    \caption{ The construction of measurement circuits for MC discrimination is demonstrated in ibm$\_$strasbourg (green), and a simulator, Qiskit Aer (yellow). Results from the simulator are shown as mean values with standard deviations. QNN-HEA and Naimark measurement ansatzes are optimized for MC measurements of ensembles (a) $S_1$ in Eq.~(\ref{eq:en1}), and (b) $S_2$ in Eq.~(\ref{eq:en2}) of two-copy states. (c) For the ensemble $S_3$ of three-copy states, QNN measurements are considered with HEA and CB. 
    } 
    \label{fig:MCM EX}
\end{figure}

To optimize a unitary circuit with a classical optimizer, we exploit Naimark and QNN measurement circuits. QNN measurement circuits may apply HEA or CB structures. We perform proof-of-principle demonstrations on a realistic, noisy device, ibm$\_$strasbourg, in which we collected data on 24/11/2024, and in the Qiskit Aer Simulator. The results are summarized in Fig.~\ref{fig:MCM EX} and Table \ref{tab:transpiled_depth_mcm}. Results from the simulator are shown as mean values with standard deviation. The confidence is maximized for the state labeled $0$ in ensembles. For ensemble $S_3$, the results obtained using block-encoding with $\delta_c=0.025$ are reported. 

A classical optimizer attempts to find optimal parameters of a Naimark measurement and fully QNN measurements with HEA or CB. Although a Naimark measurement can ultimately reach the highest confidence values, it does not achieve the confidence values on noisy quantum devices. QNN measurements with HEA, which will not construct optimal MC measurements, turn out to be effective on noisy quantum hardware. They even outperform CB for an ensemble of three-copy states.

After we optimize the maximum confidence, i.e., parameters $\Theta$ are obtained by optimizing measurement ansatzes, we can compute the probability of being inconclusive. Note that the analytic solution is known \cite{Chefles2001unambiguous}
\bea
p_{?}^{(S_1)} =0,~~
p_{?}^{(S_2)} =0.25,~~
\mathrm{and}
~~p_{?}^{(S_3)} =0.25. \nonumber
\eea
In Table~\ref{tab:inconclusive_prob_mcm}, the probabilities of inconclusive outcomes are provided. In the comparison of HEA and CB in QNN ansatzes, CB provides a lower probability of inconclusive outcomes. 

\begin{table*}[t]
\centering
\small
\caption{ Analytical and numerical inconclusive outcome probabilities for MC discrimination on ensembles $S_1$, $S_2$, and $S_3$. For ensemble $S_3$, QNN ansatzes are applied only where HEA or CB are applied. }
\label{tab:inconclusive_prob_mcm}
\renewcommand{\arraystretch}{1.2}
\setlength{\tabcolsep}{6pt}
\begin{tabular}{c|cc|cc|cc}
\hline
\textbf{Ensemble}
& \multicolumn{2}{c|}{\textbf{$S_1$}}
& \multicolumn{2}{c|}{\textbf{$S_2$}}
& \multicolumn{2}{c}{\textbf{$S_3$}}
\\
\hline
\textbf{Ansatz}
& \textbf{QNN-HEA}
& \textbf{Naimark}
& \textbf{QNN-HEA}
& \textbf{Naimark}
& \textbf{ QNN-HEA}
& \textbf{ QNN-CB}
\\
\hline
$\delta_c$
& 0.005
& 0.005
& 0.005
& 0.005
& 0.005
& 0.005
\\
\hline
\begin{tabular}[c]{@{}c@{}}
$p_{?}$\\
\end{tabular}
& 0.015
& 0.022
& 0.306
& 0.276
& 0.475
& 0.440
\\
\hline
\end{tabular}
\end{table*}

\section{Concluding remarks}
\label{sec:con}

In conclusion, we have investigated quantum measurements within the circuit model. We have constructed $l$-outcome Naimark measurements using universal gates: CNOTs together with single-qubit rotations. The construction applies to realizing arbitrary POVMs. We have shown that Naimark measurement circuits share a structural correspondence with quantum algorithms for QUBO problems, which are computationally hard. This observation suggests that a Naimark circuit is unlikely to serve as a useful ansatz for quantum optimization. As an alternative, we introduce QNN measurement circuits based on the hardware-efficient ansatz or a circuit-block layout, both of which are better suited to realistic, noisy quantum hardware: the number of CNOT gates scales linearly with system size, and entangling operations act only on neighboring qubits sequentially.

We use these constructions as the measurement ansatz: when a task of searching for optimal measurements is given, one can find optimal measurements by applying the ansatz and optimizing their parameters. We have applied these constructions to two canonical problems in optimal state discrimination: minimizing the average error probability, and maximizing the confidence of measurement outcomes. For minimum-error measurements, we have introduced a hybrid QNN scheme that retains the overall structure of a Naimark measurement while replacing the $m$-qubit universal circuit $\mathrm{U}(m)$ with a QNN circuit. For maximum-confidence measurements, we have devised a block-encoding circuit and employed QNN circuits to optimize the corresponding unitary operations.

Proof-of-principle demonstrations of both ME and MC measurements were performed on the superconducting quantum processor \texttt{ibm\_strasbourg} and on the Qiskit Aer simulator. In addition, ME discrimination has been demonstrated on the IQM Garnet. In all cases, we find that QNN-HEA is the measurement ansatz that can be efficiently optimized: the circuits converge after a relatively smaller number of training iterations. In contrast, the Naimark ansatz is the least efficient, requiring substantially more iterations to reach convergence. On the other hand, Naimark measurements can attain the optimal measurement exactly. QNN measurements may depend on the number of layers; we have shown that a QNN with $5$ layers remarkably outperforms one with $3$ layers. A clear trade-off therefore emerges between optimization efficiency and the precision with which the optimal measurement is approximated. We further note that QNN measurements are particularly well-suited to present-day quantum devices for two reasons: (i) they are efficiently optimized, and (ii) their CNOT gates respect the nearest-neighbor connectivity of typical qubit arrays.

The proposed measurement ansatzes have distinct usefulness. On the one hand, a Naimark ansatz offers advantages for achieving exact POVM elements, but optimizing its parameters may be inefficient; in particular, it contains a number of $ZZ$ interactions. On the other hand, relaxing $ZZ$ interactions with PQCs and thus composing ansatzes with QNN may lift the computational hardness. The drawback may lie in the limited expressibility of PQCs, leaving much room to improve QNN measurements. Proof-of-principle demonstrations have shown that adding layers can improve QNN measurements. We also remark that the proposed measurement ansatzes may not be biased to a particular optimizer.

We have opened a pathway to optimizing measurements on noisy intermediate-scale quantum hardware. It would be interesting to design quantum measurement circuits that are simultaneously efficient to optimize and capable of high-precision approximation on realistic, noisy hardware. In particular, minimizing resources in the framework of a Naimark measurement, deterministically and constructively, is both fundamentally and practically useful. 

\section*{Data and Software Availability }
The data and software used to support the findings of this investigation are publicly available on GitHub at [https://github.com/wonagram/Measurement-Ansatz-Naimark-QNN]. The repository contains the source code, analysis scripts, and relevant data required to reproduce the results presented in this article.

\section*{Acknowledgements}

SWY and JB were supported by the National Research Foundation of Korea (RS-2024-00408613) and the Institute for Information \& Communication Technology Promotion (IITP) (RS-2023-00229524, RS-2025-02304540, RS-2025-25464876, RS-2025-25464616). THK was supported by LG Electronics. THK extends thanks to Kevin Ferreira, Yipeng Ji, and Paria Nejat of the LG Electronics Toronto AI Lab, and
to Sean Kim of LG Electronics, AI Lab, for their support during this research.

\bibliographystyle{unsrt}
\bibliography{ref_QST}

\newpage

\newpage
\appendix

\section*{APPENDIX}
\bigskip

\section{ Algorithms for constructing a Naimark measurement }

\begin{algorithm}[h]
\caption{Construction of an $l$-outcome single-qubit POVM}
\label{alg:single_qubit_povm}
\begin{algorithmic}[1]
\Require Number of POVM outcomes $l \geq 2$
\Require System qubit $S$
\Require Binary modules $B_{S A_j}$
\Require Collective gates $CX(x)$
\Ensure Measurement circuit $U_{SA}$ implementing $l$ POVM outcomes

\State $m \gets \lceil \log_2 l \rceil$
\State Initialize ancillary qubits
\[
A_1,\ldots,A_m
\]
in the state
\[
\ket{0}^{\otimes m}.
\]

\State Apply the first binary module $B_{S A_1}$
\State $n_{\mathrm{out}} \gets 2$
\For{$L = 2,\ldots,m$}
    \State Construct the ordered set
    $\mathcal{A}(L)=\{x_1,x_2,\ldots,x_{2^{L-1}-1}\},$
    where the bit strings are ordered increasingly.
    \For{$x\in\mathcal{A}(L)$ in increasing order}   
        \State Apply $B_{S A_L}$
        \State $n_{\mathrm{out}} \gets n_{\mathrm{out}}+1$
        \If{$n_{\mathrm{out}} = 2^k$ for some integer $k$}
            \State \textbf{break}
        \EndIf
        \State Apply the collective gate $CX^{(x)}$
        \If{$n_{\mathrm{out}} = l$}
            \State \textbf{break}
        \EndIf
    \EndFor
    \If{$n_{\mathrm{out}} = l$}
        \State \textbf{break}
    \EndIf
\EndFor
\State Measure $A_1,\ldots,A_m$ in the computational basis

\For{each populated ancilla outcome $i$}
    \State Obtain 
    \[
    K_i = {}_A\!\bra{i}U_{SA}\ket{0}^{\otimes m}_A
    \]
    \State Obtain the POVM element
    \[
    M_i = K_i^\dagger K_i
    \]
\EndFor
\State \Return $U_{SA}$ and $\{M_i\}_{i=1}^{l}$
\end{algorithmic}
\end{algorithm}

\newpage

\section{Types of $CX^{(x)}$ gates}
In this section, we describe the types of $CX^{(x)}$ gates required for a Naimark measurement circuit acting on $m$ qubits. We first define the gate $\mathrm{MCX}^l_k$, which denotes a multi-controlled multi-target $X$ gate acting on $l$ target qubits with $k$ control qubits, assuming the last qubit is a control qubit. Hereafter, we use "$\mathrm{MCX}$" generically to refer to a multi-controlled multi-target $X$ gate.

\begin{figure*}
    \centering
\includegraphics[width=0.7\textwidth]{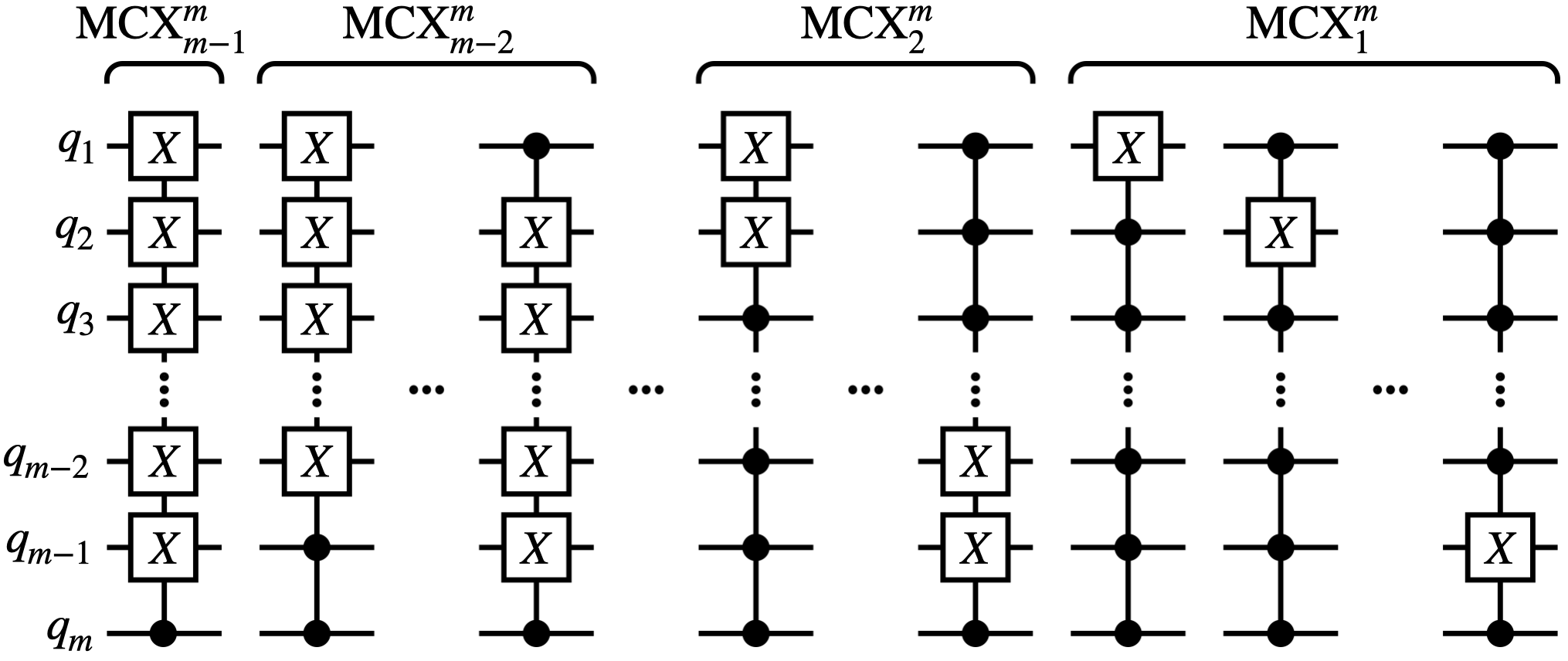}
    \caption{The figure shows the types of $CX^{(x)}$ gates required for the Naimark measurement circuit acting on $m$ qubits; in total, up to $2^{m-1}-1$ such gates can act on $m$ qubits. The $m$-bit strings $x$ are elements of the set $A(m)$. These $CX^{(x)}$ gates correspond to $\mathrm{MCX}$ gates, ranging from $\mathrm{MCX}^m_{m-1}$ to $\mathrm{MCX}^m_{1}$. They are applied in increasing order of $x$, by selecting them sequentially from left to right in the figure. For $x=x_1x_2\cdots x_m$, $x_j=0$ leads to an $X$ gate on qubit $q_j$, whereas $x_j=1$ leads to $q_j$ serving as a control in the $\mathrm{MCX}$ gate. The two-outcome POVM circuit ($m=1$) does not involve any $CX^{(x)}$ gates.}
    \label{fig: CXx}
\end{figure*}

In general, the $CX^{(x)}$ gate is a $\mathrm{MCX}$ gate that maps the register $\ket{1}^{\otimes m}$ to $\ket{x}$. In the Naimark measurement circuit, the bitstrings $\{x\}$ used for the $CX^{(x)}$ gates are elements of the set $A(m)$ (see the definition of $A(L)$ in Section~\ref{sec:single_qubit_POVM}). The $CX^{(x)}$ should be applied in increasing order of $x$, as described in the main text. This ordering is obtained by selecting gates sequentially from left to right among the $\mathrm{MCX}^m_k$ (from $k=m-1$ to $1$), as shown in Fig.\ref{fig: CXx}. Given the $m$-bit string $x=x_1x_2\cdots x_m$, $x_j=0$ corresponds to applying an $X$ gate to qubit $q_j$, whereas $x_j=1$ corresponds to using $q_j$ as a control qubit in the $\mathrm{MCX}$ gate. 

Note that the $CX^{(x)}$ gates must be applied sequentially across different $\mathrm{MCX}^m_k$ gates, although arbitrary ordering among gates within a single $\mathrm{MCX}^m_k$ gate is allowed. Otherwise, swaps occur among the pre-existing ancillary registers through $B_{SA}$ gates. In total, the number of $CX^{(x)}$ gates for $m$ ancillary qubits is given by $2^{m-1}-1$.

\section{CNOT Count in a Multi-qubit POVM Circuit}
\label{sec:appcnot}

In this section, we derive the total number of CNOTs, $\mathcal{O}(2^n l^2)$, required for the hybrid Naimark-QNN measurement circuit, as shown in Fig.~\ref{fig:Yordan multi qubit PQC}.

\subsection{CNOT count in the $i$-th $CX^{(x)}$ gate} \label{sec: num. CXx}
We analyze the CNOT count required to implement the $i$-th $CX^{(x)}$ gate. Both the $i$-th $B_{SA}$ gate and the subsequent $CX^{(x)}$ gate act on $\lfloor \log_2 i \rfloor +1$ ancillary qubits. Here, the $i$-th $CX^{(x)}$ gate denotes the one applied immediately after the $i$-th $B_{SA}$ gate.

\begin{figure}[ht]
    \centering
    \includegraphics[width=0.24\textwidth]{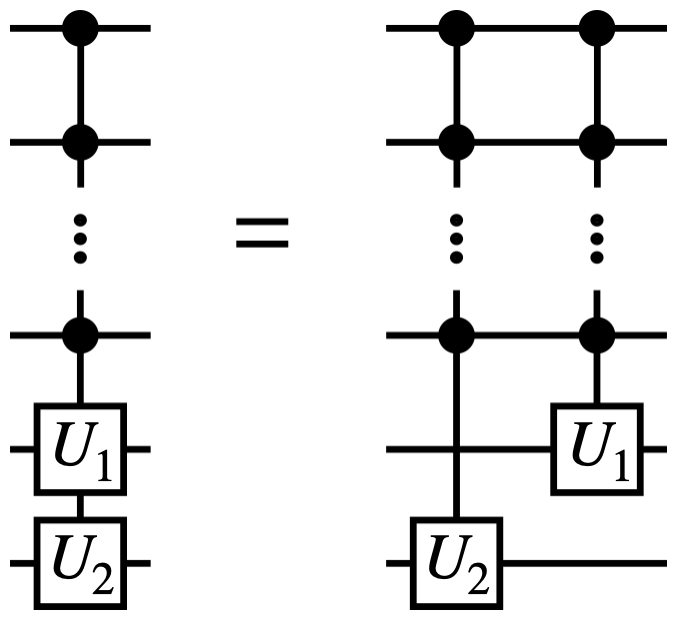}
    \caption{Decomposition of multi-controlled $U_1$ and $U_2$ gates, where both unitaries are applied separately and controlled by the same set of qubits.}
    \label{fig: U decomposition}
\end{figure}

Assume that the bitstring $x$ contains $k$ bits equal to $0$, corresponding to $k$ $X$ gates applied to the corresponding qubits. The $CX^{(x)}$ has $\lfloor \log_2 i \rfloor +1-k$ control qubits. Using the decomposition scheme described in Fig.~\ref{fig: U decomposition}, the gate is decomposed into $k$ multi-controlled $X$ gates, each having $\lfloor \log_2 i \rfloor +1-k$ control qubits.
According to Ref.~\cite{PhysRevA.52.3457}, a multi-controlled $X$ gate with $m$ control qubits (for $m \geq 5$) can be decomposed into \(8(m-3)\) three-qubit Toffoli gates. Each three-qubit Toffoli can, in turn, be decomposed into 6 CNOT gates~\cite{shende2009on}. Thus, the multi-controlled $X$ gate with $m$ control qubits can be decomposed into $48(m-3)$ CNOTs. Therefore, the total number of CNOT gates is given by

\bea
48k(\lfloor \log_2 i \rfloor-k-2)
\eea
where $1 \leq k \leq \lfloor \log_2 i \rfloor$. This expression attains its maximum when $k=\frac{\lfloor \log_2 i \rfloor-2}{2}$, yielding a maximum number of CNOT gates: 

\bea
48\bigg(\frac{\lfloor \log_2 i \rfloor-2}{2}\bigg)^2.
\eea

By straightforward algebra, we obtain:

\bea
\mathcal{O}(48k(\lfloor \log_2 i \rfloor-k-2)) = \mathcal{O}((\log_2 i)^2).
\eea

\subsection{CNOT count in the uniformly controlled rotation gate at the \(i\)-th $B_{SA}$ gate}
The uniformly controlled rotation gate in the \(i\)-th $B_{SA}$ gate has two types of controls: $n$ system qubits that act as uniformly controlling qubits, and $\lfloor \log_2 i \rfloor$ ancillary qubits that act as standard control qubits.

\begin{figure*}
    \centering
    \includegraphics[width=0.6\textwidth]{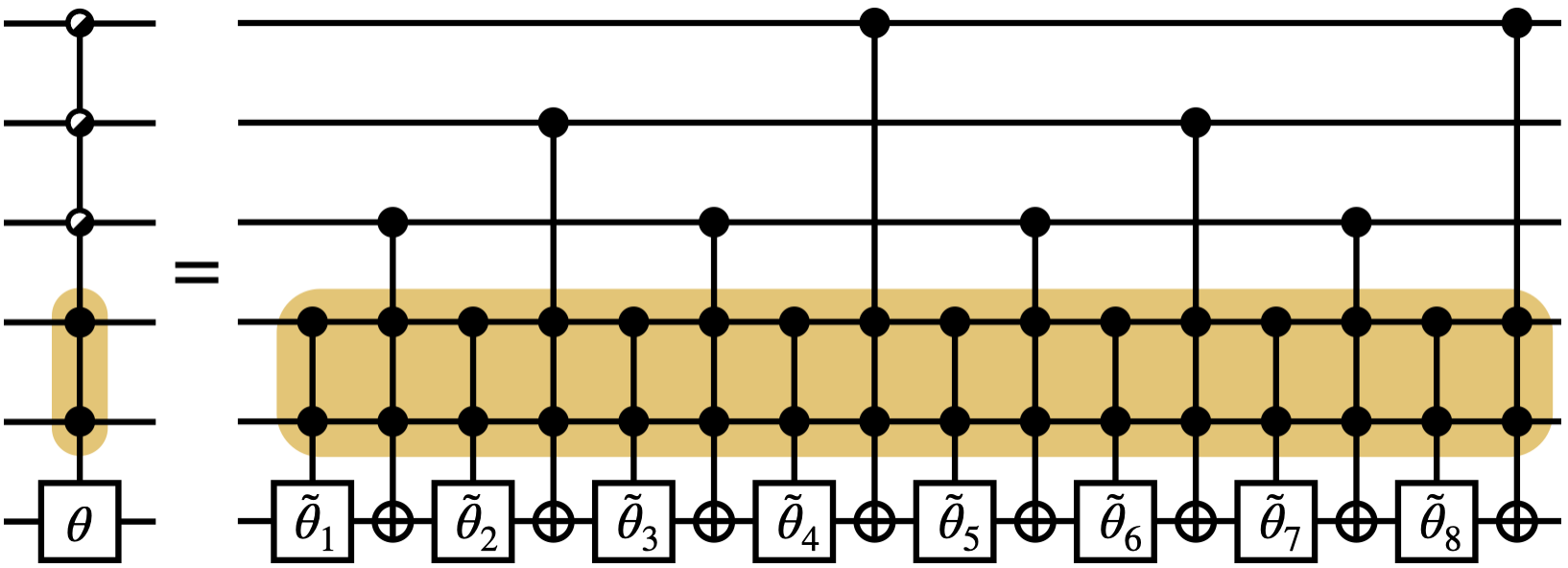}
    \caption{Decomposition scheme of a uniformly controlled rotation gate with three uniformly controlling qubits and two additional control qubits. While the original decomposition of the uniformly controlled gate follows the method of~\cite{Mottonen2004quantum}, this figure illustrates how the gate structure extends when extra control qubits are included.}
    \label{fig: Uni con gate decomp}
\end{figure*}

As shown in Ref.~\cite{Mottonen2004quantum}, a uniformly controlled rotation gate with $m$ uniformly controlling qubits is decomposed into \(2^m\) single-qubit rotations and $2^m$ CNOT gates. Thus, using the decomposition scheme in Fig.~\ref{fig: Uni con gate decomp}, our gate is decomposed into $2^n$ multi-controlled rotation gates with $\lfloor \log_2 i \rfloor$ control qubits. Furthermore, the multi-controlled rotation gate with $m$ control qubits can be decomposed into $2^m -2$ CNOT gates and $2m$ single-qubit rotation gates~\cite{POVM_qc_2019}. Therefore, the total number of CNOT gates is given by 

\bea
2^n(2^{\lfloor \log_2 i \rfloor}-2). 
\eea

\subsection{CNOT count in the $V$ gate at the \(i\)-th $B_{SA}$ gate}
A $B_{SA}$ gate contains two $V$ gates, each composed of $n$ single-qubit rotation gates and $n-1$ CNOT gates. Each $V$ gate is controlled by $\lfloor \log_2 i \rfloor+1$ qubits. We first consider the rotation gates, more precisely, multi-controlled rotation gates. Since a multi-controlled rotation gate with $m$ control qubits is decomposed into $2^m-2$ CNOT gates, the total number of CNOTs is given by 

\bea
2n(2^{\lfloor \log_2 i \rfloor+1}-2)
\eea

where the factor of 2 accounts for the two $V$ gates in the $B_{SA}$ gate. 

We proceed to analyze the CNOT gates in the $V$ gates, more precisely, $n-1$ multi-controlled $X$ gates with $\lfloor \log_2 i \rfloor+2$ control qubits. From the previous discussion, there are a total of 

\bea
96(n-1)(\lfloor \log_2 i \rfloor-1)
\eea
CNOTs, again including a factor of 2 for the two $V$ gates in the $B_{SA}$ gate.

Therefore, at the $i$-th $B_{SA}$ gate and the $CX^{(x)}$ gate, the uniformly controlled rotation gates dominate the CNOT count.

\bea
O(2^n(2^{\lfloor \log_2 i \rfloor}-2))=\mathcal{O}(2^n i)
\eea

The total number of CNOT gates for an $l$-outcome measurement with $n$ ancilla qubits can be obtained by summing over all $B_{SA}$ gates and reads
\bea
\sum_{i=1}^{l-1}O(2^n i ) = \mathcal{O}(2^n l^2)
\eea
A particular instance for a single-qubit POVM reproduces $\mathcal{O}(l^2)$~\cite{POVM_qc_2019}. The count above can be seen as a generalization to the multi-qubit POVM.

\section{Total CNOT Count in the MCM Circuit}
In this section, we analyze the CNOT count required to implement the circuit in Fig.~\ref{fig:bee} where $U_S$ is treated as an HEA. We consider two cases: 
\begin{itemize}
    \item{{\it The case $l\leq2^n$}: The circuit does not require any ancilla qubits.}
    \item{{\it The case $l > 2^n$}: The circuit requires $n_A=\lceil \log_2 l\rceil-n$ ancillary qubits, excluding Ancilla$1$ and Ancilla$2$ shown in Fig.~\ref{fig: MCM BE}.}
\end{itemize}
After applying the BE gate on the system qubits, $U_S$ acts on $n$ qubits when $l\leq2^n$, and $n+n_A$ qubits otherwise; hence, in general, $U_S$ acts on $m=\max(n,\lceil \log_2l\rceil)$ qubits. The $O_K$ in the BE gate shown in Fig.~\ref{fig:bee} contains \(4m^2\) CNOT gates by construction. In addition, each SWAP gate is decomposed into three CNOT gates, contributing $3m$ CNOTs. Furthermore, the HEA with $L$ layers in $U_S$ has $L(m-1)$ CNOTs. Therefore, the CNOT count in the MCM circuit is $\mathcal{O}(m^2)$.

\section{Detailed analytical and numerical results of ME discrimination}
\subsection{Analytical optimal measurements}

The optimal POVM elements for ensemble $S_1$ are given by
\begin{align}
&M_0 = 0,\nonumber\\
&M_1 = |+\rangle\langle+|, \nonumber\\
&M_2 = |-\rangle\langle-|. \nonumber
\end{align}
Since the states in ensemble $S_2$ are symmetrical and equiprobable, the optimal POVM elements are given by the square-root measurement~\cite{Ban1997optimum}. The corresponding measurement vectors and POVM elements are
\begin{align}
|\mu_i\rangle=\frac{1}{\sqrt{3}}\rho^{-\frac{1}{2}}|\psi_i\rangle,\quad M_i^{\mathrm{opt}}=|\mu_i\rangle\langle\mu_i|.\nonumber
\end{align}
The measurement vectors are explicitly given by
\begin{align}
|\mu_0\rangle &= \frac{\sqrt{3}(\sqrt{2}+2)}{6}|00\rangle+\frac{\sqrt{3}(\sqrt{2}-2)}{6}|11\rangle,\nonumber\\
|\mu_1\rangle &= \frac{\sqrt{6}-\sqrt{3}}{6}|00\rangle+\frac{1}{2}|01\rangle+\frac{1}{2}|10\rangle+\frac{\sqrt{6}+\sqrt{3}}{6}|11\rangle ,\nonumber\\
|\mu_2\rangle &= \frac{\sqrt{6}-\sqrt{3}}{6}|00\rangle-\frac{1}{2}|01\rangle-\frac{1}{2}|10\rangle+\frac{\sqrt{6}+\sqrt{3}}{6}|11\rangle.\nonumber
\end{align}
Likewise, the optimal POVM elements of $S_3$ are given by the square-root measurement. 
\begin{align}
|\mu_i\rangle=\frac{1}{\sqrt{3}}\rho^{-\frac{1}{2}}|\psi_i\rangle,\quad M_i^{\mathrm{opt}}=|\mu_i\rangle\langle\mu_i|.\nonumber
\end{align}
The measurement vectors are
\begin{align}
|\mu_0\rangle &=\frac{\sqrt{3}(1+\sqrt{6})}{6}|000\rangle\nonumber\\
&\quad+\frac{\sqrt{2}-\sqrt{3}}{6}\big(|011\rangle+|101\rangle+|110\rangle\big) \nonumber\\
|\mu_1\rangle &=\frac{\sqrt{3}(\sqrt{6}-2)}{12}|000\rangle+\frac{\sqrt{6}}{12}\big(|001\rangle+|010\rangle+|100\rangle\big) \nonumber\\
&\quad+\frac{\sqrt{3}(\sqrt{6}+6)}{36}\big(|011\rangle+|101\rangle+|110\rangle\big)+\frac{\sqrt{6}}{4} |111\rangle \nonumber\\
|\mu_2\rangle &=\frac{\sqrt{3}(\sqrt{6}-2)}{12}|000\rangle-\frac{\sqrt{6}}{12}\big(|001\rangle+|010\rangle+|100\rangle\big) \nonumber\\
&\quad+\frac{\sqrt{3}(\sqrt{6}+6)}{36}\big(|011\rangle+|101\rangle+|110\rangle\big)-\frac{\sqrt{6}}{4} |111\rangle \nonumber
\end{align}

\subsection{POVM elements obtained from the simulator}\label{sec:POVMobtained}

Among thirty independent optimization runs, we report the result corresponding to the maximum guessing probability.

We evaluate the purity of POVM elements, 
\bea
\mathrm{purity}(M_i) = \mathrm{tr}\Big[\big(\frac{M_i}{\mathrm{tr}[M_i]}\big)^2\Big]
\eea
to quantify their rank. Note that, as shown in the previous subsection, optimal POVM elements are rank-one.

Firstly, for the ensemble $S_1$, the optimized POVM elements from the Naimark measurement are given below: 
\begin{align}
&M_1 =\nonumber\\
&\footnotesize\begin{pmatrix}
4.984\times 10^{-1} & 4.993\times 10^{-1} + 6.835\times 10^{-3}\mathrm{i} \\
4.993\times 10^{-1} - 6.835\times 10^{-3}\mathrm{i} & 5.003\times 10^{-1}
\end{pmatrix}\nonumber\\
&M_2 =\nonumber\\
&\footnotesize\begin{pmatrix}
5.008\times 10^{-1} & -4.998\times 10^{-1} - 6.837\times 10^{-3}\mathrm{i} \\
-4.998\times 10^{-1} + 6.837\times 10^{-3}\mathrm{i} & 4.990\times 10^{-1}
\end{pmatrix}\nonumber
\end{align}
which show 
\bea
\mathrm{purity}(M_1) = \mathrm{purity}(M_2)
 =1.000.\nonumber
 \eea
We recall that $M_0=0$ from the analytic solution, meaning no-measurement is optimal and thus the confidence follows only from the {\it a priori} probability.

Secondly, for the ensemble $S_2$, the optimized POVM elements from the Naimark measurement are given below: 
\begin{widetext}
\begin{align}
&M_0 = \nonumber\\
&\scriptsize\begin{pmatrix}
9.643\times 10^{-1} & -7.199\times 10^{-3} - 6.680\times 10^{-3}\mathrm{i} & 1.423\times 10^{-2} - 1.921\times 10^{-2}\mathrm{i} & -1.760\times 10^{-1} + 1.063\times 10^{-2}\mathrm{i} \\
-7.199\times 10^{-3} + 6.680\times 10^{-3}\mathrm{i} & 4.933\times 10^{-1} & -4.763\times 10^{-1} - 4.920\times 10^{-2}\mathrm{i} & -1.634\times 10^{-2} - 2.404\times 10^{-2}\mathrm{i} \\
1.423\times 10^{-2} + 1.921\times 10^{-2}\mathrm{i} & -4.763\times 10^{-1} + 4.920\times 10^{-2}\mathrm{i} & 4.720\times 10^{-1} & 1.765\times 10^{-2} + 1.734\times 10^{-2}\mathrm{i} \\
-1.760\times 10^{-1} - 1.063\times 10^{-2}\mathrm{i} & -1.634\times 10^{-2} + 2.404\times 10^{-2}\mathrm{i} & 1.765\times 10^{-2} - 1.734\times 10^{-2}\mathrm{i} & 3.505\times 10^{-2}
\end{pmatrix}\nonumber\\
&M_1 = \nonumber\\
&\scriptsize\begin{pmatrix}
1.523\times 10^{-2} & 6.377\times 10^{-2} - 3.927\times 10^{-3}\mathrm{i} & 5.766\times 10^{-2} + 5.122\times 10^{-3}\mathrm{i} & 8.118\times 10^{-2} + 2.352\times 10^{-3}\mathrm{i} \\
6.377\times 10^{-2} + 3.927\times 10^{-3}\mathrm{i} & 2.744\times 10^{-1} & 2.531\times 10^{-1} + 3.112\times 10^{-2}\mathrm{i} & 3.490\times 10^{-1} + 2.650\times 10^{-2}\mathrm{i} \\
5.766\times 10^{-2} - 5.122\times 10^{-3}\mathrm{i} & 2.531\times 10^{-1} - 3.112\times 10^{-2}\mathrm{i} & 2.556\times 10^{-1} & 3.370\times 10^{-1} - 1.809\times 10^{-2}\mathrm{i} \\
8.118\times 10^{-2} - 2.352\times 10^{-3}\mathrm{i} & 3.490\times 10^{-1} - 2.650\times 10^{-2}\mathrm{i} & 3.370\times 10^{-1} + 1.809\times 10^{-2}\mathrm{i} & 4.577\times 10^{-1}
\end{pmatrix}\nonumber\\
&M_2 = \nonumber\\
&\scriptsize\begin{pmatrix}
2.044\times 10^{-2} & -5.657\times 10^{-2} + 1.061\times 10^{-2}\mathrm{i} & -7.189\times 10^{-2} + 1.409\times 10^{-2}\mathrm{i} & 9.484\times 10^{-2} - 1.299\times 10^{-2}\mathrm{i} \\
-5.657\times 10^{-2} - 1.061\times 10^{-2}\mathrm{i} & 2.323\times 10^{-1} & 2.231\times 10^{-1} + 1.808\times 10^{-2}\mathrm{i} & -3.327\times 10^{-1} - 2.459\times 10^{-3}\mathrm{i} \\
-7.189\times 10^{-2} - 1.409\times 10^{-2}\mathrm{i} & 2.231\times 10^{-1} - 1.808\times 10^{-2}\mathrm{i} & 2.724\times 10^{-1} & -3.547\times 10^{-1} + 7.461\times 10^{-4}\mathrm{i} \\
9.484\times 10^{-2} + 1.299\times 10^{-2}\mathrm{i} & -3.327\times 10^{-1} + 2.459\times 10^{-3}\mathrm{i} & -3.547\times 10^{-1} - 7.461\times 10^{-4}\mathrm{i} & 5.073\times 10^{-1}\nonumber
\end{pmatrix}
\end{align}
\end{widetext}
Then, we have purities,
\bea
\mathrm{purity}~(M_0) = 0.498,~ 
\mathrm{purity}~(M_1)= 0.976, \nonumber\\
\mathrm{purity}~(M_2) = 0.933. \nonumber
\eea
The resulting POVM elements are close to rank-one operators but $M_0$.

For the ensemble $S_3$, the optimized POVM elements from the Hybrid QNN measurement are given below:
\begin{widetext}
\begin{align}
&\operatorname{Re}(M_0) =\nonumber\\
&\scriptsize\begin{pmatrix}
10.000\times 10^{-1} & 0 & 0 & 1.210\times 10^{-5} & 0 & -1.166\times 10^{-5} & 7.796\times 10^{-6} & 0 \\
0 & 1.969\times 10^{-1} & -1.210\times 10^{-5} & 0 & 1.166\times 10^{-5} & 0 & 0 & 1.642\times 10^{-6} \\
0 & -1.210\times 10^{-5} & 8.015\times 10^{-2} & 0 & -7.796\times 10^{-6} & 0 & 0 & -1.132\times 10^{-6} \\
1.210\times 10^{-5} & 0 & 0 & 1.296\times 10^{-3} & 0 & -1.642\times 10^{-6} & 1.132\times 10^{-6} & 0 \\
0 & 1.166\times 10^{-5} & -7.796\times 10^{-6} & 0 & 3.115\times 10^{-5} & 0 & 0 & -4.070\times 10^{-8} \\
-1.166\times 10^{-5} & 0 & 0 & -1.642\times 10^{-6} & 0 & 1.715\times 10^{-3} & 4.070\times 10^{-8} & 0 \\
7.796\times 10^{-6} & 0 & 0 & 1.132\times 10^{-6} & 0 & 4.070\times 10^{-8} & 7.596\times 10^{-4} & 0 \\
0 & 1.642\times 10^{-6} & -1.132\times 10^{-6} & 0 & -4.070\times 10^{-8} & 0 & 0 & 7.613\times 10^{-6}
\end{pmatrix}\nonumber\\
&\operatorname{Im}(M_0) = \nonumber\\
&\scriptsize\begin{pmatrix}
0 & -4.294\times 10^{-3} & 2.875\times 10^{-3} & 0 & -2.710\times 10^{-3} & 0 & 0 & -3.291\times 10^{-8} \\
4.294\times 10^{-3} & 0 & 0 & 6.113\times 10^{-4} & 0 & -5.289\times 10^{-4} & 3.291\times 10^{-8} & 0 \\
-2.875\times 10^{-3} & 0 & 0 & -4.217\times 10^{-4} & 0 & 3.291\times 10^{-8} & -2.151\times 10^{-4} & 0 \\
0 & -6.113\times 10^{-4} & 4.217\times 10^{-4} & 0 & -3.291\times 10^{-8} & 0 & 0 & -3.490\times 10^{-6} \\
2.710\times 10^{-3} & 0 & 0 & 3.291\times 10^{-8} & 0 & 9.003\times 10^{-6} & -2.277\times 10^{-6} & 0 \\
0 & 5.289\times 10^{-4} & -3.291\times 10^{-8} & 0 & -9.003\times 10^{-6} & 0 & 0 & 5.335\times 10^{-6} \\
0 & -3.291\times 10^{-8} & 2.151\times 10^{-4} & 0 & 2.277\times 10^{-6} & 0 & 0 & -4.021\times 10^{-6} \\
3.291\times 10^{-8} & 0 & 0 & 3.490\times 10^{-6} & 0 & -5.335\times 10^{-6} & 4.021\times 10^{-6} & 0
\end{pmatrix}\nonumber\\
&\operatorname{Re}(M_1) =\nonumber\\
&\scriptsize\begin{pmatrix}
1.949\times 10^{-5} & 1.353\times 10^{-4} & -4.422\times 10^{-5} & 6.055\times 10^{-6} & -4.498\times 10^{-5} & 1.133\times 10^{-4} & -3.116\times 10^{-5} & -2.585\times 10^{-6} \\
1.353\times 10^{-4} & 3.970\times 10^{-1} & 8.564\times 10^{-4} & 1.255\times 10^{-1} & 1.562\times 10^{-3} & 1.248\times 10^{-1} & -3.614\times 10^{-2} & 1.310\times 10^{-3} \\
-4.422\times 10^{-5} & 8.564\times 10^{-4} & 4.571\times 10^{-1} & -3.780\times 10^{-2} & 6.620\times 10^{-3} & 4.329\times 10^{-1} & -1.337\times 10^{-1} & 1.536\times 10^{-3} \\
6.055\times 10^{-6} & 1.255\times 10^{-1} & -3.780\times 10^{-2} & 4.971\times 10^{-1} & 4.528\times 10^{-1} & -2.991\times 10^{-3} & 1.221\times 10^{-3} & 1.393\times 10^{-1} \\
-4.498\times 10^{-5} & 1.562\times 10^{-3} & 6.620\times 10^{-3} & 4.528\times 10^{-1} & 4.969\times 10^{-1} & -4.052\times 10^{-2} & -1.398\times 10^{-1} & 1.015\times 10^{-3} \\
1.133\times 10^{-4} & 1.248\times 10^{-1} & 4.329\times 10^{-1} & -2.991\times 10^{-3} & -4.052\times 10^{-2} & 4.970\times 10^{-1} & 1.818\times 10^{-3} & 1.397\times 10^{-1} \\
-3.116\times 10^{-5} & -3.614\times 10^{-2} & -1.337\times 10^{-1} & 1.221\times 10^{-3} & -1.398\times 10^{-1} & 1.818\times 10^{-3} & 5.037\times 10^{-1} & 4.541\times 10^{-1} \\
-2.585\times 10^{-6} & 1.310\times 10^{-3} & 1.536\times 10^{-3} & 1.393\times 10^{-1} & 1.015\times 10^{-3} & 1.397\times 10^{-1} & 4.541\times 10^{-1} & 4.940\times 10^{-1}
\end{pmatrix}\nonumber\\
&\operatorname{Im}(M_1) =\nonumber\\
&\scriptsize\begin{pmatrix}
0 & 2.130\times 10^{-3} & -1.405\times 10^{-3} & 2.017\times 10^{-3} & 1.334\times 10^{-3} & -7.954\times 10^{-4} & -1.543\times 10^{-4} & 4.609\times 10^{-6} \\
-2.130\times 10^{-3} & 0 & 6.604\times 10^{-4} & -4.549\times 10^{-3} & -2.368\times 10^{-3} & 5.421\times 10^{-4} & 2.625\times 10^{-3} & 1.456\times 10^{-4} \\
1.405\times 10^{-3} & -6.604\times 10^{-4} & 0 & -1.339\times 10^{-3} & -4.607\times 10^{-4} & 3.328\times 10^{-2} & -1.411\times 10^{-4} & 4.219\times 10^{-3} \\
-2.017\times 10^{-3} & 4.549\times 10^{-3} & 1.339\times 10^{-3} & 0 & -2.373\times 10^{-3} & -1.227\times 10^{-4} & 1.357\times 10^{-3} & 2.782\times 10^{-4} \\
-1.334\times 10^{-3} & 2.368\times 10^{-3} & 4.607\times 10^{-4} & 2.373\times 10^{-3} & 0 & -1.673\times 10^{-3} & 4.665\times 10^{-3} & -1.272\times 10^{-3} \\
7.954\times 10^{-4} & -5.421\times 10^{-4} & -3.328\times 10^{-2} & 1.227\times 10^{-4} & 1.673\times 10^{-3} & 0 & -4.463\times 10^{-3} & -4.681\times 10^{-3} \\
1.543\times 10^{-4} & -2.625\times 10^{-3} & 1.411\times 10^{-4} & -1.357\times 10^{-3} & -4.665\times 10^{-3} & 4.463\times 10^{-3} & 0 & 1.862\times 10^{-2} \\
-4.609\times 10^{-6} & -1.456\times 10^{-4} & -4.219\times 10^{-3} & -2.782\times 10^{-4} & 1.272\times 10^{-3} & 4.681\times 10^{-3} & -1.862\times 10^{-2} & 0
\end{pmatrix}\nonumber\\
&\operatorname{Re}(M_2) = \nonumber\\
&\scriptsize\begin{pmatrix}
1.993\times 10^{-5} & -1.353\times 10^{-4} & 4.422\times 10^{-5} & -1.816\times 10^{-5} & 4.498\times 10^{-5} & -1.016\times 10^{-4} & 2.337\times 10^{-5} & 2.585\times 10^{-6} \\
-1.353\times 10^{-4} & 4.061\times 10^{-1} & -8.443\times 10^{-4} & -1.255\times 10^{-1} & -1.573\times 10^{-3} & -1.248\times 10^{-1} & 3.614\times 10^{-2} & -1.312\times 10^{-3} \\
4.422\times 10^{-5} & -8.443\times 10^{-4} & 4.628\times 10^{-1} & 3.780\times 10^{-2} & -6.612\times 10^{-3} & -4.329\times 10^{-1} & 1.337\times 10^{-1} & -1.535\times 10^{-3} \\
-1.816\times 10^{-5} & -1.255\times 10^{-1} & 3.780\times 10^{-2} & 5.016\times 10^{-1} & -4.528\times 10^{-1} & 2.993\times 10^{-3} & -1.222\times 10^{-3} & -1.393\times 10^{-1} \\
4.498\times 10^{-5} & -1.573\times 10^{-3} & -6.612\times 10^{-3} & -4.528\times 10^{-1} & 5.031\times 10^{-1} & 4.052\times 10^{-2} & 1.398\times 10^{-1} & -1.015\times 10^{-3} \\
-1.016\times 10^{-4} & -1.248\times 10^{-1} & -4.329\times 10^{-1} & 2.993\times 10^{-3} & 4.052\times 10^{-2} & 5.013\times 10^{-1} & -1.818\times 10^{-3} & -1.397\times 10^{-1} \\
2.337\times 10^{-5} & 3.614\times 10^{-2} & 1.337\times 10^{-1} & -1.222\times 10^{-3} & 1.398\times 10^{-1} & -1.818\times 10^{-3} & 4.956\times 10^{-1} & -4.541\times 10^{-1} \\
2.585\times 10^{-6} & -1.312\times 10^{-3} & -1.535\times 10^{-3} & -1.393\times 10^{-1} & -1.015\times 10^{-3} & -1.397\times 10^{-1} & -4.541\times 10^{-1} & 5.059\times 10^{-1}
\end{pmatrix}\nonumber\\
&\operatorname{Im}(M_2) =\nonumber\\
&\scriptsize\begin{pmatrix}
0 & 2.164\times 10^{-3} & -1.470\times 10^{-3} & -2.017\times 10^{-3} & 1.376\times 10^{-3} & 7.954\times 10^{-4} & 1.543\times 10^{-4} & -4.576\times 10^{-6} \\
-2.164\times 10^{-3} & 0 & -6.604\times 10^{-4} & 3.937\times 10^{-3} & 2.368\times 10^{-3} & -1.315\times 10^{-5} & -2.625\times 10^{-3} & -1.456\times 10^{-4} \\
1.470\times 10^{-3} & 6.604\times 10^{-4} & 0 & 1.761\times 10^{-3} & 4.607\times 10^{-4} & -3.328\times 10^{-2} & 3.562\times 10^{-4} & -4.219\times 10^{-3} \\
2.017\times 10^{-3} & -3.937\times 10^{-3} & -1.761\times 10^{-3} & 0 & 2.373\times 10^{-3} & 1.227\times 10^{-4} & -1.357\times 10^{-3} & -2.747\times 10^{-4} \\
-1.376\times 10^{-3} & -2.368\times 10^{-3} & -4.607\times 10^{-4} & -2.373\times 10^{-3} & 0 & 1.664\times 10^{-3} & -4.663\times 10^{-3} & 1.272\times 10^{-3} \\
-7.954\times 10^{-4} & 1.315\times 10^{-5} & 3.328\times 10^{-2} & -1.227\times 10^{-4} & -1.664\times 10^{-3} & 0 & 4.463\times 10^{-3} & 4.676\times 10^{-3} \\
-1.543\times 10^{-4} & 2.625\times 10^{-3} & -3.562\times 10^{-4} & 1.357\times 10^{-3} & 4.663\times 10^{-3} & -4.463\times 10^{-3} & 0 & -1.862\times 10^{-2} \\
4.576\times 10^{-6} & 1.456\times 10^{-4} & 4.219\times 10^{-3} & 2.747\times 10^{-4} & -1.272\times 10^{-3} & -4.676\times 10^{-3} & 1.862\times 10^{-2} & 0 
\end{pmatrix} \nonumber
\end{align}
\end{widetext}
Then, we have purities,
\bea
\mathrm{purity}~(M_0) = 0.637,~ 
\mathrm{purity}~(M_1)= 0.271,~
\mathrm{and} \nonumber\\
\mathrm{purity}~(M_2) = 0.269. \nonumber
\eea
The resulting POVM elements are noisy being far from rank-one operators.


\section{Analytical optimal POVM elements for maximum-confidence discrimination}\label{sec:MC_optimal}

For pure states, the analytical optimal POVM elements for maximum-confidence discrimination, up to the coefficients $c_i$ are given by $M_i=c_i\rho^{-1}|\psi_i\rangle\langle\psi|\rho^{-1}$~\cite{MCM2006}. If $\rho$ is not full rank, we replace $\rho^{-1}$ with the Moore--Penrose pseudoinverse. Here, we report the analytical forms of $\rho^{-1}|\psi\rangle_i$ for each state. 

For $S_1$, the $\rho^{-1}|\psi_i\rangle$ are given by 
\bea
\rho^{-1}|\psi\rangle_0 &=& \frac{9}{7}|0\rangle, \nonumber\\
\rho^{-1}|\psi\rangle_1 &=& \frac{3\sqrt{6}}{7}|0\rangle+\frac{3\sqrt{3}}{2}|1\rangle, \nonumber\\
\rho^{-1}|\psi\rangle_2 &=& \frac{3\sqrt{6}}{7}|0\rangle-\frac{3\sqrt{3}}{2}|1\rangle. \nonumber
\eea

For $S_2$, the $\rho^{-1}|\psi_i\rangle$ are given by 
\bea
\rho^{-1}|\psi\rangle_0 &=& 3|00\rangle-|11\rangle, \nonumber\\
\rho^{-1}|\psi\rangle_1 &=& \sqrt{3}|01\rangle+\sqrt{3}|10\rangle+2|11\rangle, \nonumber\\
\rho^{-1}|\psi\rangle_2 &=& -\sqrt{3}|01\rangle-\sqrt{3}|10\rangle+2|11\rangle. \nonumber
\eea

For $S_3$, the $\rho^{-1}|\psi_i\rangle$ are given by 
\bea
\rho^{-1}|\psi\rangle_0 &=& 3|000\rangle-\frac{1}{3}\big(|011\rangle+|101\rangle+|110\rangle\big), \nonumber\\
\rho^{-1}|\psi\rangle_1 &=& \frac{1}{\sqrt{3}}\big(|001\rangle+|010\rangle+|100\rangle\big)\nonumber\\
&&+\frac{4}{3}\big(|011\rangle+|101\rangle+|110\rangle\big)+\sqrt{3}|111\rangle, \nonumber\\
\rho^{-1}|\psi\rangle_2 &=& -\frac{1}{\sqrt{3}}\big(|001\rangle+|010\rangle+|100\rangle\big)\nonumber\\
&&+\frac{4}{3}\big(|011\rangle+|101\rangle+|110\rangle\big)-\sqrt{3}|111\rangle, \nonumber
\eea

\end{document}